\documentclass[prx,twocolumn,preprintnumbers,amsmath,amssymb,dvipdfmx]{revtex4}

\usepackage{graphicx}
\usepackage{bm}
\usepackage{color}
\usepackage{here}

\begin{document}


\title{Dynamical stability of two-dimensional metals in the periodic table}
\author{Shota Ono}
\email{shota\_o@gifu-u.ac.jp}
\affiliation{Department of Electrical, Electronic and Computer Engineering, Gifu University, Gifu 501-1193, Japan}

\begin{abstract}
We study the dynamical stability of elemental two-dimensional (2D) metals from Li to Pb by calculating the phonon band structure from first principles, where 2D structures are assumed to be planer hexagonal, buckled honeycomb, and buckled square lattice structures. We show the relationship between the stability of 2D structures and that of three-dimensional structures. This provides a material design concept for alloys, where the similarity with regard to the stable 2D structures, rather than the energetic stability of alloy, is important to yield dynamically stable alloys. 
\end{abstract}

\maketitle

Most elemental metals have either the face-centered cubic (FCC), hexagonal closed-packed (HCP), or body-centered cubic (BCC) structures as its ground state \cite{AM}. Some elemental metals in a different structure can be dynamically stable, i.e., metastable \cite{grimvall}, which are of interest both from a fundamental and practical point of view \cite{togo,schonecker,huang,yin,kong}. In this paper, we explore dynamically stable structures in two-dimensional (2D) metals and apply them as a building block for computational materials synthesis.

Graphene (C), atomically thin carbon layers, has honeycomb (HC) structure with zero thickness \cite{NG}. In contrast, other 2D elements in group 14 have buckled HC (bHC) as a dynamically stable structure \cite{cahangirov,balendhran}. For example, silicene (Si), germanene (Ge), and stanene (Sn) have a total thickness of 0.45, 0.69, and 0.85 \AA, respectively \cite{balendhran}. This is attributed to the fact that the bonding in the bulk prefers $sp^3$ to $sp^2$ character, producing the diamond structure but not the graphite-like structure. Thus far, noble metals (Cu \cite{yang3}, Ag \cite{yang2}, and Au \cite{yang1}), borophene (B) \cite{borophene1,borophene2} and gallenene (Ga) \cite{kochat} in group 13, phosphorene (P) \cite{BP}, arsenene (As) \cite{AsSb}, antimonene (Sb) \cite{AsSb}, and bithmuthene (Bi) \cite{akturk} in group 15, poloniumene (Po) in group 16 \cite{ono2020} have been reported as stable 2D materials theoretically and/or experimentally, while B and Ga require external strains (or substrate) to be stabilized. Recently, the stability of elemental 2D metals have been investigated by calculating the total energy within the density-functional theory (DFT) \cite{nevalaita,hwang}. Although the planer hexagonal (HX) \cite{nevalaita} and buckled structures \cite{hwang} have been proposed as stable structures, no phonon band structure calculations have been performed. 

\begin{figure*}[t]
\center
\includegraphics[scale=0.55]{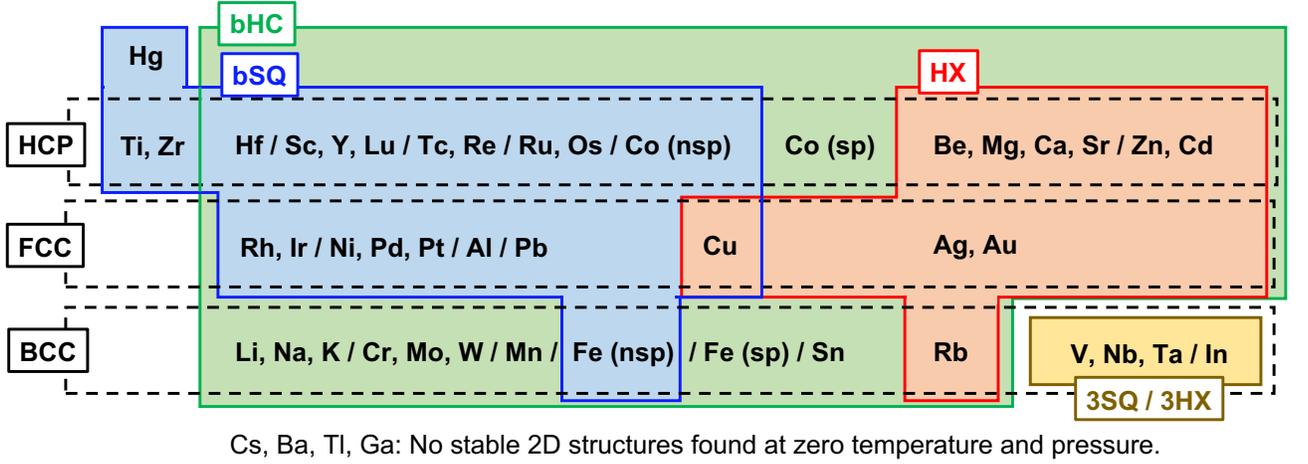}
\caption{Relationship of the dynamical stability between 2D structures (HX, bHC, and bSQ) and 3D structures (HCP, FCC, and BCC) in elemental metals. The stability of multilayered structures (3SQ and 3HX) is investigated if the HX, bHC, and bSQ structures are unstable. The stability property of spin-polarized (sp) Co and Fe is different from that of non-spin-polarized (nsp) ones. Stable crystal structures in 3D metals are extracted from the periodic table depicted in Ref.~\cite{AM}. The cubic Mn and tetragonal In and Sn are included to the group of BCC metals. Hg has the rhombohedral structure as its ground state \cite{AM}. Note that 2D Ga becomes dynamically stable when strains are imposed \cite{kochat}. } \label{fig1} 
\end{figure*}

The aim of this paper is to study the dynamical stability of 2D metals in the periodic table from Li to Pb by assuming the planar HX, bHC, and buckled square (bSQ) structures. Figure \ref{fig1} shows how the stability of 2D metals is related to that of three-dimensional (3D) metals: (i) If the HX structure is dynamically stable in an elemental metal, the HCP and/or FCC structures are also stable (HX$\rightarrow$HCP and/or FCC) except Rb; (ii) If the bHC structure is dynamically stable only, the BCC structure is stable (bHC$\rightarrow$BCC): (iii) If the bSQ structure is dynamically stable only, the HCP structure is stable (bSQ$\rightarrow$HCP) except Hg; and (iv) If the bHC and bSQ structures are dynamically stable, where the bSQ is the metastable phase, the HCP (FCC) structure is stable for group 3, 4, 7, and 8 (group 9 and 10) metals (bHC and bSQ$\rightarrow$HCP or FCC). The rest of the paper will be devoted to show this result, to discuss the magnetic effects, and to provide an application to the material design for dynamically stable alloys.  

\begin{figure}[tt]
\center
\includegraphics[scale=0.35]{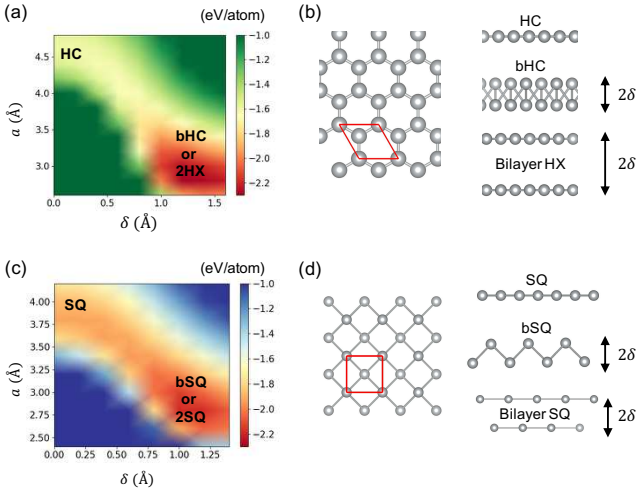}
\caption{(a) The total energy variation as a function of $a$ and $\delta$ and (b) top and side views for the bHC (or 2HX) structures of Ag. (c) and (d): Same as (a) and (b), respectively, but for the bSQ (or 2SQ) structure. The unit cell is enclosed by red lines. The layer thickness (or the interlayer distance in the bilayer structure) is equal to $2\delta$. } \label{fig2} 
\end{figure}

\begin{figure}[t]	
\center
\includegraphics[scale=0.30]{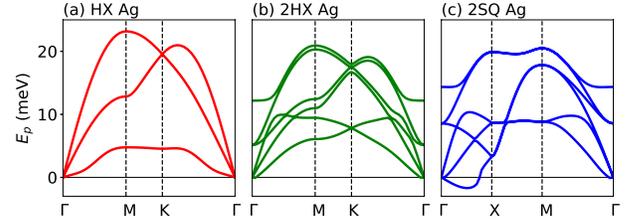}
\caption{Phonon band structure of Ag in (a) HX, (b) 2HX, and (c) 2SQ structures.  } \label{fig3} 
\end{figure}


First, we show how the buckled structures become energetically more stable than the planar structure. To demonstrate it, we calculate the total energy of 2D metals based on DFT implemented in \texttt{Quantum ESPRESSO} (\texttt{QE}) code \cite{qe}. Spin-polarized calculations are performed for Cr, Mn, Fe, Co, and Ni. The unit cell of the bHC structure includes two atoms, whose positions are given by $(0, 0, +\delta)$ and $(0, a/\sqrt{3},-\delta)$ with the lattice constant $a$ and the buckling height $\delta$ (see Fig.~\ref{fig2}(b)). If the size of $\delta$ is large enough to satisfy the inequality $(a/\sqrt{3})^2 + (2\delta)^2 > a^2$, the bHC can be regarded as the bilayer HX (2HX) structure. Figure \ref{fig2}(a) shows the total energy variation as a function of $\delta$ and $a$ for 2D Ag, where the total energy is minimum when $\delta = 0.0$ \AA \ and $a\simeq 4.6$ \AA \ and when $\delta \simeq 1.2$ \AA \ and $a\simeq 2.8$ \AA. The local minimum of the former and the latter corresponds to the planar HC and 2HX structures, respectively. The 2HX is more stable than the planar HC and HX structures by 0.716 and 0.164 eV/atom, respectively. We also consider the bSQ structure with two atoms included in a unit cell, where the positions of atoms are given by $(0, 0, +\delta)$ and $(a/2, a/2,-\delta)$, as shown in Fig.~\ref{fig2}(d). The bSQ becomes the bilayer SQ (2SQ) structure when the inequality $(a/\sqrt{2})^2 + (2\delta)^2 > a^2$ is satisfied. We plot the total energy variation as a function of $\delta$ and $a$ for 2D Ag in Fig.~\ref{fig1}(c). The two local minimum correspond to the planar SQ and the 2SQ structures. The values of $a$ giving the minimum total energy are close to the nearest-neighbor (NN) interatomic distance in FCC Ag (2.89 \AA). By setting the initial lattice parameter to the NN interatomic distance in 3D metals \cite{kittel}, we optimize the geometry of the HX, bSQ (or 2SQ), and bHC (or 2HX) structures from Li to Pb. More computational details, the cohesive energy, and the structural parameters of these structures are provided in Appendix. 

Next, we study the dynamical stability of the optimized geometries by performing the phonon band structure calculations within the density-functional perturbation theory (DFPT) \cite{dfpt} implemented in \texttt{QE} code \cite{qe}. The phonon energy is defined by $E_p = {\rm sgn}(\omega^2) \hbar\vert\omega\vert$, where $\rm {sgn}$ is the sign function, $\hbar$ is the Planck constant, and $\omega$ is the phonon frequency obtained by diagonalizing the dynamical matrix. $E_p$ is negative when $\omega$ is an imaginary number. 

Figure \ref{fig3} shows the phonon band structure of Ag. The HX and 2HX structures are dynamically stable: No imaginary frequencies are observed. The former case is consistent with the previous calculations \cite{yang2}. However, the 2SQ structure is unstable because the imaginary frequency appears along $\Gamma$-X line. Note that the 2SQ is energetically more stable than the planar HX structure by 0.057 eV/atom. This implies that the large cohesive energy does not always lead to the dynamical stability, as has been observed in many alloys \cite{sun}, while it can lead to an increase in the flexural phonon energy particularly at the point M. The computational details and the phonon band structures of other elements are provided in Appendix. Below, we will not distinguish the bHC (bSQ) and 2HX (2SQ) structures and use the ``bHC'' and ``bSQ'' only. 

{\it Stable 2D structures.} We begin with Zn and Cd that have the HX structure. It is well known that these have HCP structure, in which the ratio of the lattice constants ($c/a=1.86$ and $1.89$ for Zn and Cd, respectively) is much larger than the ideal value $c/a=1.63$ \cite{AM}. It is reasonable that they can be exfoliated easily and have monolayer HX structure. The alkali earth (Be, Mg, Ca, and Sr) and noble metals (Cu, Ag, and Au) that are classified into divalent and monovalent metals having two and one $s$ electrons in the outermost shell, respectively. Since the distribution of the $s$ electron is spherically symmetric around the nucleus, the densely packed HX structure with no buckling is realized. In this sense, it is strange that the HX structure is unstable in alkali metals (Li, Na, and K). 

All transition metals have buckled or multilayered structures. For these metals, the electronic states around the Fermi level consist of a mixture of $s$ and $p \ (d)$ electrons. Such a hybridized orbital is less spherical symmetry, creating the buckled structures, as realized in group 14 semiconductors (Si and Ge) \cite{cahangirov}. With respect to the stability, the trivalent metal of Al is similar to transition metals rather than free-electron metals (Na and K). It is interesting to note that in Cr, Hg, Mo, Nb, Ti, V, W, and Zr, the cohesive energy of bSQ is larger than that of bHC, although the energetic stability is not necessary to produce the dynamical stability of the bSQ structure. 

In case that the HX, bHC, and bSQ structures are unstable, we examine the dynamical stability of the trilayer structures. The BCC V, Nb, and Ta are dynamically stable when trilayer SQ (3SQ) structure is assumed, whereas the tetragonal In is dynamically stable in trilayer HX (3HX) structure, where the positions of atoms in a unit cell are (0, 0, 0), and ($a/2$, $a/2$, $\pm 2\delta$) for 3SQ and (0, 0, 0), and ($0$, $a/\sqrt{3}$, $\pm 2\delta$) for 3HX structures. Unfortunately, no dynamically stable structures are found for Cs, Ba, Ga, and Tl at zero pressure. The instability of these elements are due to the instability against the out-of-plane (ZA) vibrations. Fortunately, the imaginary frequencies appear around $\Gamma$ point only, meaning that these 2D metals would be stable when the lateral size of the sample is smaller than the wavelength of the corresponding ZA phonons. 
 
\begin{figure}[ttt]
\center
\includegraphics[scale=0.42]{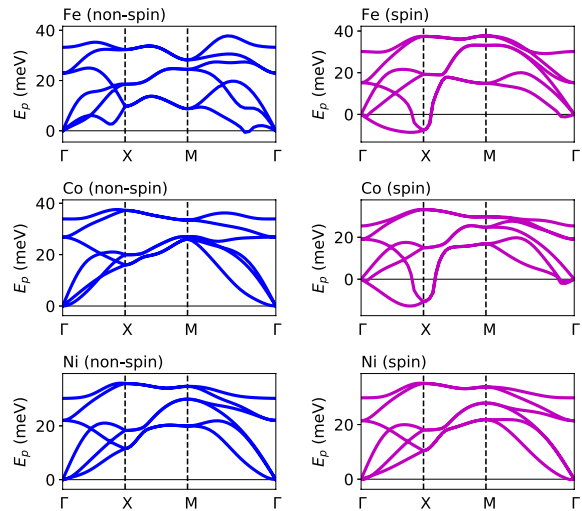}
\caption{The phonon band structure of Fe, Co, and Ni that have bSQ structure. Left for non-spin-polarized and right for spin-polarized calculations. }\label{fig4}
\end{figure}

{\it Magnetic effects.} Through analyses of the magnetic moment in period-4 metals, we find that the behavior of bHC is quite analogous to that of HCP phase. The bHC phase of Cr, Mn, and Fe has nonmagnetic, while that of Co and Ni has ferromagnetic state with the magnetic moment 1.89 and 0.82 $\mu_{\rm B}$ per atom, respectively (the Bohr magneton $\mu_{\rm B}$). For 3D case, the HCP phase of Mn and Fe has nonmagnetic, while that of Cr, Co, and Ni has ferromagnetic state with 0.12, 1.60, and 0.59 $\mu_{\rm B}$ per atom, respectively \cite{podgorny}. Within non-spin-polarized calculations, bHC Co and Ni are also dynamically stable. This is reasonable because HCP Co and Ni are elastically stable in the absence of magnetic effects \cite{guo1}. 

The trend of the dynamical stability in bSQ is different from that of bHC structure. Figure \ref{fig4} shows the phonon band structure of bSQ Fe, Co, and Ni: Left and right for the non-spin-polarized and spin-polarized calculations, respectively. Within non-spin-polarized calculations, the bSQ structure is dynamically stable. However, except Ni, those structures become dynamically unstable when magnetic effects are included, where the absolute magnetic moment per atom is 2.77, 1.94, and 0.82 $\mu_{B}$ for ferromagnetic Fe, Co, and Ni, respectively. This trend is similar to the stability of FCC Fe, where it is elastically stable and unstable in nonmagnetic and ferromagnetic states, respectively \cite{guo1}. 

{\it Material design principles.} We discuss how the dynamical stability of the 2D structure is related to that of the 3D structure in elemental metals. The property (i), i.e., HX$\rightarrow$HCP and/or FCC, can be derived from the fact that the FCC and HCP structures consist of HX monolayers stacked along the (111) direction and the $c$ axis, respectively \cite{AM}. The property (ii), i.e., bHC or trilayer$\rightarrow$BCC, is also a reasonable conclusion because the BCC structure is obtained by stacking the bHC and 3SQ structures along the (111) direction and $c$ axis, respectively. The property (iii), i.e., bSQ$\rightarrow$HCP, is derived from the fact that the HCP structure is obtained by stacking the elongated bSQ (along the diagonal direction of the square shaped unit cell) along the $c$ axis. The property (iv), i.e., bHC and bSQ$\rightarrow$HCP or FCC, may be understood as in the properties of (i) and (iii), while it is difficult to disentangle the stability properties of FCC and HCP. The understanding of the metastability of bSQ phase will be a key to resolve this issue. 

\begin{figure}[ttt]
\center
\includegraphics[scale=0.38]{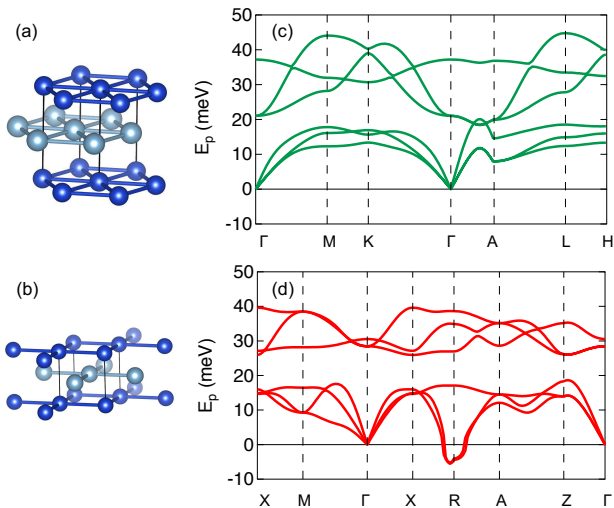}
\caption{Schematic illustration of (a) B$_h$ and (b) L1$_0$ structures. The phonon band structure of (c) B$_h$ and (d) L1$_0$ AlCu alloys. }\label{fig5} 
\end{figure}

We propose that the stability relationship between 2D and 3D metals can be applied to a prediction of the dynamical stability of 3D alloys. In the present study, we focus on the binary Al-Cu system that has a complex phase diagram \cite{wit}. For the case of Al$_{0.5}$Cu$_{0.5}$, it is natural to consider that AlCu has the B$_h$ (WC) structure with HX Al and HX Cu layers stacked alternately, as shown in Fig.~\ref{fig5}(a), since the bHC and HX structures in Al and Cu are dynamically stable, respectively, as shown in Fig.~\ref{fig1}. Meanwhile, the L1$_0$ (CuAu) structure (see Fig.~\ref{fig5}(b)) has been predicted to be more energetically stable structure \cite{zhou_alloy}. This may be reasonable because Al and Cu have the bSQ as a metastable structure and also because the L1$_0$ structure consists of the SQ lattice of Al and the displaced SQ lattice of Cu stacked alternately along $c$ axis. Figures \ref{fig5}(c) and \ref{fig5}(d) show the phonon band structure of AlCu alloy in the B$_h$ and L1$_0$ structures, respectively, indicating that the B$_h$ is dynamically stable, whereas the L1$_0$ is unstable. The stability of the B$_h$ phase is robust against the thermal fluctuations up to 2500 K, which is confirmed by performing {\it ab initio} molecular dynamics (MD) simulations implemented in \texttt{QE} code \cite{qe}; see Appendix for the computational details. We again emphasize that the energetic stability is not enough to yield the dynamical stability in alloys; For the present case, the L1$_0$ structure is more stable than the B$_h$ structure by 0.058 eV per unit cell. We can demonstrate that B$_h$ CuZn is dynamically stable, while CuZn has B2 (CsCl) structure as its ground state \cite{beck}, as provided in Appendix. Our approach based on the list of dynamically stable structures in Fig.~\ref{fig1} provides a novel principle for computational design of alloys.

{\it Future perspective.} In order to establish a design principle for alloys, it would be important to understand the origin of metastability in both elemental metals and alloys \cite{togo,sun} and to calculate the energy landscape as a function of atom positions \cite{gaston} in alloys. It is interesting to discuss how the present approach is related to the Pettifor's approach for the stability of binary alloys \cite{pettifor}. The present study is limited to the phonons of free-standing and ideal thin films of elemental metals at zero temperature and pressure. It would be interesting to study how the effects of substrate (or strain) \cite{kochat}, chemical adsorption \cite{hwang}, vacancies \cite{nevalaita2}, grain boundaries \cite{diery}, and temperature \cite{hellman} influence the phonon band structure and the dynamical stability of 2D metals. We note that the effect of spin-orbit coupling is important in heavy elements because it strongly influences the dynamical stability of 2D Bi \cite{akturk} and 2D Po \cite{ono2020} as well as the metastability of FCC Pt in the HCP structure \cite{schonecker}. For fundamental interest, it is interesting to calculate the electron-phonon coupling function (Eliashberg function) of 2D metals in order to find novel 2D superconductors \cite{zhang2015,liu2018,yao,yankowitz}. It is also interesting to explore whether 2D Mn has more complex structures, since 3D Mn has a BCC-related structure with 58 atoms per unit cell \cite{hobbs}.  

\begin{acknowledgments}
This study is supported by the Nikki-Saneyoshi Foundation. A part of numerical calculations has been done using the facilities of the Supercomputer Center, the Institute for Solid State Physics, the University of Tokyo.
\end{acknowledgments}

\appendix 

\begin{figure*}[t] 
\center
\includegraphics[scale=0.50]{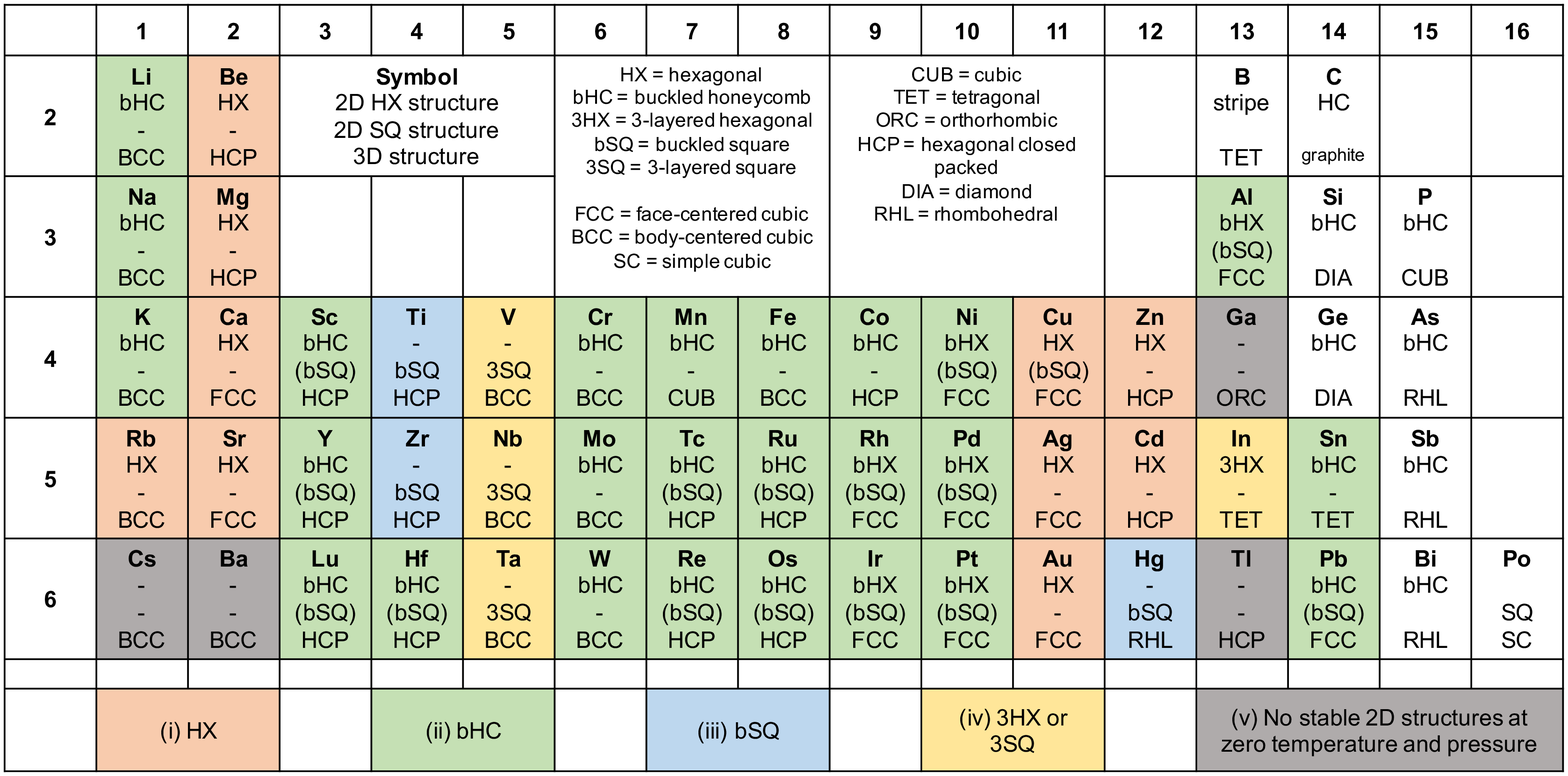}
\caption{Periodic table for elemental 2D materials. The present results are colored. The stable 2D structures are indicated (The structure in parenthesis indicates the metastable phase). Stable crystal structures in 3D materials are from Ref.~\cite{AM}. The previously investigated structures that have dynamical stability include semiconductors of C \cite{NG}, Si, Ge, and Sn \cite{balendhran}, P \cite{BP}, and As and Sb \cite{AsSb}, and metals of Cu \cite{yang3}, Ag \cite{yang2}, Au \cite{yang1}, B \cite{borophene1,borophene2}, Ga \cite{kochat}, Bi \cite{akturk}, and Po \cite{ono2020}.  }
\label{fig_S1} 
\end{figure*}

\section{Computational details}
We calculate the total energy of 2D metals based on DFT implemented in \texttt{Quantum ESPRESSO} (\texttt{QE}) code \cite{qe}. The effects of exchange and correlation are treated within GGA-PBE \cite{pbe} and GGA-PBEsol \cite{pbesol}. We use the ultrasoft pseudopotentials generated by the scheme of Ref.~\cite{dalcorso}, i.e., \texttt{pslibrary.1.0.0}. For Na, Mn, and Tc, \texttt{pslibrary.0.2}, \texttt{pslibrary.0.3.1}, and \texttt{pslibrary.0.3.0} are used, respectively. The cutoff energies for the wavefunction and the charge density are 80 Ry and 800 Ry, respectively, which are well above the suggested cutoff energies. The self-consistent calculations are performed by using 30$\times$30$\times$1 $k$ grid for 2D metals and 15$\times$15$\times$15 $k$ grid for 3D alloys \cite{MK}. The primitive lattice vectors in units of the lattice constant $a$ are $(1,0,0)$, $(-1/2,\sqrt{3}/2,0)$, and $(0,0,c/a)$ for HX, 2HX, and 3HX structures, and $(1,0,0)$, $(0,1,0)$, and $(0,0,c/a)$ for 2SQ and 3SQ structures. The size of $c$ is fixed to be 14 \AA. The Marzari-Vanderbilt smearing \cite{smearing} with a broadening of $\sigma=0.02$ Ry is used for all calculations.

The phonon band structure calculations are performed within the density-functional perturbation theory \cite{dfpt} implemented in \texttt{QE} code \cite{qe}. For HX and bHC structures, 8$\times$8$\times$1 $q$ grid (10 $q$ points) is used, while for bSQ structure, 6$\times$6$\times$1 $q$ grid (10 $q$ points) is used. For 3HX and 3SQ structures, 6$\times$6$\times$1 $q$ grid (7 and 10 $q$ points, respectively) is used. In cases of bHC W, bSQ W, and anti-ferromagnetic phase of HX Cr and Mn, the size of $q$ grid is decreased to 6$\times$6$\times$1, 4$\times$4$\times$1, and 4$\times$2$\times$1, respectively, in order to reduce the computational costs. For AlCu CuZn alloys, 4$\times$4$\times$4 $q$ grid is used. 

The {\it ab initio} molecular dynamics (MD) simulations for B$_h$ AlCu and CuZn alloys are performed using \texttt{QE} code \cite{qe}. The ionic temperature is controlled via the velocity scaling and increased from 500 K up to 3000 K with an increment of 500 K. A 2$\times$2$\times$2 supercell with 16 atoms is considered. The Newton's equation is integrated by using Verlet algorithm with a time step of 0.967 fs (20 a.u.) and up to 2.0 ps (2100 MD steps).   

Throughout the main text and this supplemental material, we discuss the dynamical stability of 2D metals and 3D alloys based on the GGA-PBE calculation results. It would be valuable to note that the dynamically stable structures obtained from the GGA-PBEsol calculations are almost the same as those obtained from the GGA-PBE calculations. The differences are as follows: Within GGA-PBEsol, HX K, bSQ Hg, and 3HX Tl are dynamically stable, while bSQ Cu and HX Au are unstable. We have not studied the dynamical stability of Nb and Tc because no GGA-PBEsol potentials are available.

\begin{table*}
\begin{center}
\caption{$E_{{\rm c}, j}$ (eV/atom), $a_j$ (\AA), and $\delta_j$ (\AA) for $j=1,2,3$. ``nsp'' and ``sp'' in the parenthesis indicate the non-spin-polarized and spin-polarized calculations, respectively.}
{
\begin{tabular}{lcccccccc}\hline
   & $E_{{\rm c}, 1}$ \hspace{5mm}  & $a_1$ \hspace{5mm}   &   $E_{{\rm c}, 2}$ \hspace{5mm}  &  $a_2$ \hspace{5mm}  & $\delta_2$  \hspace{5mm}   &   $E_{{\rm c}, 3}$ \hspace{5mm}  &  $a_3$ \hspace{5mm}  & $\delta_3$    \\ \hline
Ag	\hspace{5mm} &	2.144 	\hspace{5mm} &	2.794 	\hspace{5mm} &	2.307 	\hspace{5mm} &	2.855 	\hspace{5mm} &	1.239 	\hspace{5mm} &	2.201 	\hspace{5mm} &	2.802 	\hspace{5mm} &	1.082 	\\
Al	\hspace{5mm} &	2.842 	\hspace{5mm} &	2.682 	\hspace{5mm} &	3.217 	\hspace{5mm} &	2.748 	\hspace{5mm} &	1.227 	\hspace{5mm} &	3.085 	\hspace{5mm} &	2.713 	\hspace{5mm} &	1.088 	\\
Au	\hspace{5mm} &	2.845 	\hspace{5mm} &	2.748 	\hspace{5mm} &	2.897 	\hspace{5mm} &	2.773 	\hspace{5mm} &	1.442 	\hspace{5mm} &	2.722 	\hspace{5mm} &	2.756 	\hspace{5mm} &	1.143 	\\
Ba	\hspace{5mm} &	1.203 	\hspace{5mm} &	4.467 	\hspace{5mm} &	1.593 	\hspace{5mm} &	4.493 	\hspace{5mm} &	1.763 	\hspace{5mm} &	1.547 	\hspace{5mm} &	4.339 	\hspace{5mm} &	1.609 	\\
Be	\hspace{5mm} &	2.997 	\hspace{5mm} &	2.126 	\hspace{5mm} &	3.354 	\hspace{5mm} &	2.157 	\hspace{5mm} &	0.959 	\hspace{5mm} &	3.181 	\hspace{5mm} &	2.090 	\hspace{5mm} &	0.799 	\\
Ca	\hspace{5mm} &	1.181 	\hspace{5mm} &	3.866 	\hspace{5mm} &	1.584 	\hspace{5mm} &	3.924 	\hspace{5mm} &	1.503 	\hspace{5mm} &	1.530 	\hspace{5mm} &	3.792 	\hspace{5mm} &	1.340 	\\
Cd	\hspace{5mm} &	0.487 	\hspace{5mm} &	2.922 	\hspace{5mm} &	0.642 	\hspace{5mm} &	2.950 	\hspace{5mm} &	1.489 	\hspace{5mm} &	0.505 	\hspace{5mm} &	2.938 	\hspace{5mm} &	1.378 	\\
Co (nsp)	\hspace{5mm} &	5.028 	\hspace{5mm} &	2.302 	\hspace{5mm} &	5.902 	\hspace{5mm} &	2.379 	\hspace{5mm} &	0.975 	\hspace{5mm} &	5.619 	\hspace{5mm} &	2.366 	\hspace{5mm} &	0.821 	\\
Co (sp)	\hspace{5mm} &	5.463 	\hspace{5mm} &	2.355 	\hspace{5mm} &	6.178 	\hspace{5mm} &	2.441 	\hspace{5mm} &	0.974 	\hspace{5mm} &	5.901 	\hspace{5mm} &	2.397 	\hspace{5mm} &	0.874 	\\
Cr (nsp)	\hspace{5mm} &	6.440 	\hspace{5mm} &	2.340 	\hspace{5mm} &	7.388 	\hspace{5mm} &	2.788 	\hspace{5mm} &	0.716 	\hspace{5mm} &	7.411 	\hspace{5mm} &	2.280 	\hspace{5mm} &	1.053 	\\
Cr (sp)	\hspace{5mm} &	6.735 	\hspace{5mm} &	2.690 	\hspace{5mm} &	7.388 	\hspace{5mm} &	2.788 	\hspace{5mm} &	0.716 	\hspace{5mm} &	7.411 	\hspace{5mm} &	2.280 	\hspace{5mm} &	1.053 	\\
Cs	\hspace{5mm} &	0.554 	\hspace{5mm} &	5.385 	\hspace{5mm} &	0.635 	\hspace{5mm} &	5.213 	\hspace{5mm} &	2.110 	\hspace{5mm} &	0.626 	\hspace{5mm} &	5.250 	\hspace{5mm} &	2.031 	\\
Cu	\hspace{5mm} &	3.154 	\hspace{5mm} &	2.428 	\hspace{5mm} &	3.407 	\hspace{5mm} &	2.496 	\hspace{5mm} &	1.063 	\hspace{5mm} &	3.280 	\hspace{5mm} &	2.468 	\hspace{5mm} &	0.912 	\\
Fe (nsp)	\hspace{5mm} &	5.780 	\hspace{5mm} &	2.287 	\hspace{5mm} &	6.903 	\hspace{5mm} &	2.400 	\hspace{5mm} &	0.927 	\hspace{5mm} &	6.496 	\hspace{5mm} &	2.377 	\hspace{5mm} &	0.800 	\\
Fe (sp)	\hspace{5mm} &	6.314 	\hspace{5mm} &	2.405 	\hspace{5mm} &	6.903 	\hspace{5mm} &	2.400 	\hspace{5mm} &	0.927 	\hspace{5mm} &	6.893 	\hspace{5mm} &	2.391 	\hspace{5mm} &	0.969 	\\
Ga	\hspace{5mm} &	2.261 	\hspace{5mm} &	2.749 	\hspace{5mm} &	2.516 	\hspace{5mm} &	2.822 	\hspace{5mm} &	1.298 	\hspace{5mm} &	2.497 	\hspace{5mm} &	2.683 	\hspace{5mm} &	1.225 	\\
Hf	\hspace{5mm} &	4.850 	\hspace{5mm} &	2.919 	\hspace{5mm} &	6.091 	\hspace{5mm} &	3.152 	\hspace{5mm} &	1.140 	\hspace{5mm} &	6.082 	\hspace{5mm} &	3.051 	\hspace{5mm} &	0.999 	\\
Hg	\hspace{5mm} &	0.093 	\hspace{5mm} &	3.534 	\hspace{5mm} &	0.128 	\hspace{5mm} &	3.553 	\hspace{5mm} &	1.471 	\hspace{5mm} &	0.123 	\hspace{5mm} &	3.574 	\hspace{5mm} &	1.249 	\\
In	\hspace{5mm} &	1.912 	\hspace{5mm} &	3.162 	\hspace{5mm} &	2.165 	\hspace{5mm} &	3.226 	\hspace{5mm} &	1.459 	\hspace{5mm} &	2.136 	\hspace{5mm} &	3.069 	\hspace{5mm} &	1.382 	\\
Ir	\hspace{5mm} &	6.948 	\hspace{5mm} &	2.564 	\hspace{5mm} &	7.739 	\hspace{5mm} &	2.662 	\hspace{5mm} &	1.101 	\hspace{5mm} &	7.312 	\hspace{5mm} &	2.663 	\hspace{5mm} &	0.908 	\\
K 	\hspace{5mm} &	0.700 	\hspace{5mm} &	4.527 	\hspace{5mm} &	0.810 	\hspace{5mm} &	4.636 	\hspace{5mm} &	1.901 	\hspace{5mm} &	0.787 	\hspace{5mm} &	4.430 	\hspace{5mm} &	1.763 	\\
Li	\hspace{5mm} &	1.311 	\hspace{5mm} &	3.091 	\hspace{5mm} &	1.561 	\hspace{5mm} &	3.102 	\hspace{5mm} &	1.159 	\hspace{5mm} &	1.538 	\hspace{5mm} &	2.960 	\hspace{5mm} &	1.085 	\\
Lu	\hspace{5mm} &	2.595 	\hspace{5mm} &	3.338 	\hspace{5mm} &	3.533 	\hspace{5mm} &	3.433 	\hspace{5mm} &	1.320 	\hspace{5mm} &	3.434 	\hspace{5mm} &	3.419 	\hspace{5mm} &	1.103 	\\
Mg	\hspace{5mm} &	0.952 	\hspace{5mm} &	3.058 	\hspace{5mm} &	1.241 	\hspace{5mm} &	3.087 	\hspace{5mm} &	1.354 	\hspace{5mm} &	1.106 	\hspace{5mm} &	3.046 	\hspace{5mm} &	1.174 	\\
Mn (nsp)	\hspace{5mm} &	6.322 	\hspace{5mm} &	2.296 	\hspace{5mm} &	7.437 	\hspace{5mm} &	2.447 	\hspace{5mm} &	0.921 	\hspace{5mm} &	7.119 	\hspace{5mm} &	2.618 	\hspace{5mm} &	0.604 	\\
Mn (sp)	\hspace{5mm} &	6.871 	\hspace{5mm} &	2.573 	\hspace{5mm} &	7.437 	\hspace{5mm} &	2.447 	\hspace{5mm} &	0.921 	\hspace{5mm} &	7.119 	\hspace{5mm} &	2.618 	\hspace{5mm} &	0.604 	\\
Mo	\hspace{5mm} &	7.931 	\hspace{5mm} &	2.597 	\hspace{5mm} &	8.879 	\hspace{5mm} &	3.102 	\hspace{5mm} &	0.808 	\hspace{5mm} &	8.907 	\hspace{5mm} &	2.553 	\hspace{5mm} &	1.138 	\\
Na	\hspace{5mm} &	0.937 	\hspace{5mm} &	3.659 	\hspace{5mm} &	1.064 	\hspace{5mm} &	3.681 	\hspace{5mm} &	1.519 	\hspace{5mm} &	1.036 	\hspace{5mm} &	3.519 	\hspace{5mm} &	1.404 	\\
Nb	\hspace{5mm} &	7.122 	\hspace{5mm} &	2.700 	\hspace{5mm} &	7.865 	\hspace{5mm} &	3.187 	\hspace{5mm} &	0.907 	\hspace{5mm} &	8.157 	\hspace{5mm} &	2.663 	\hspace{5mm} &	1.179 	\\
Ni (nsp)	\hspace{5mm} &	4.164 	\hspace{5mm} &	2.338 	\hspace{5mm} &	4.769 	\hspace{5mm} &	2.408 	\hspace{5mm} &	1.020 	\hspace{5mm} &	4.573 	\hspace{5mm} &	2.377 	\hspace{5mm} &	0.874 	\\
Ni (sp)	\hspace{5mm} &	4.253 	\hspace{5mm} &	2.355 	\hspace{5mm} &	4.827 	\hspace{5mm} &	2.425 	\hspace{5mm} &	1.010 	\hspace{5mm} &	4.614 	\hspace{5mm} &	2.397 	\hspace{5mm} &	0.867 	\\
Os	\hspace{5mm} &	8.202 	\hspace{5mm} &	2.553 	\hspace{5mm} &	9.510 	\hspace{5mm} &	2.681 	\hspace{5mm} &	1.034 	\hspace{5mm} &	8.861 	\hspace{5mm} &	2.698 	\hspace{5mm} &	0.850 	\\
Pb	\hspace{5mm} &	2.795 	\hspace{5mm} &	3.305 	\hspace{5mm} &	3.139 	\hspace{5mm} &	3.535 	\hspace{5mm} &	1.322 	\hspace{5mm} &	3.129 	\hspace{5mm} &	3.428 	\hspace{5mm} &	1.195 	\\
Pd	\hspace{5mm} &	2.847 	\hspace{5mm} &	2.631 	\hspace{5mm} &	3.407 	\hspace{5mm} &	2.701 	\hspace{5mm} &	1.175 	\hspace{5mm} &	3.238 	\hspace{5mm} &	2.658 	\hspace{5mm} &	1.024 	\\
Pt	\hspace{5mm} &	5.082 	\hspace{5mm} &	2.611 	\hspace{5mm} &	5.391 	\hspace{5mm} &	2.695 	\hspace{5mm} &	1.205 	\hspace{5mm} &	5.150 	\hspace{5mm} &	2.650 	\hspace{5mm} &	1.040 	\\
Rb	\hspace{5mm} &	0.619 	\hspace{5mm} &	4.925 	\hspace{5mm} &	0.715 	\hspace{5mm} &	4.974 	\hspace{5mm} &	2.051 	\hspace{5mm} &	0.692 	\hspace{5mm} &	4.754 	\hspace{5mm} &	1.903 	\\
Re	\hspace{5mm} &	8.633 	\hspace{5mm} &	2.573 	\hspace{5mm} &	10.057 	\hspace{5mm} &	2.724 	\hspace{5mm} &	1.033 	\hspace{5mm} &	9.498 	\hspace{5mm} &	2.944 	\hspace{5mm} &	0.672 	\\
Rh	\hspace{5mm} &	4.643 	\hspace{5mm} &	2.551 	\hspace{5mm} &	5.569 	\hspace{5mm} &	2.634 	\hspace{5mm} &	1.092 	\hspace{5mm} &	5.265 	\hspace{5mm} &	2.620 	\hspace{5mm} &	0.923 	\\
Ru	\hspace{5mm} &	6.321 	\hspace{5mm} &	2.528 	\hspace{5mm} &	7.592 	\hspace{5mm} &	2.663 	\hspace{5mm} &	1.015 	\hspace{5mm} &	7.147 	\hspace{5mm} &	2.668 	\hspace{5mm} &	0.845 	\\
Sc	\hspace{5mm} &	2.706 	\hspace{5mm} &	3.149 	\hspace{5mm} &	3.619 	\hspace{5mm} &	3.283 	\hspace{5mm} &	1.199 	\hspace{5mm} &	3.553 	\hspace{5mm} &	3.218 	\hspace{5mm} &	1.038 	\\
Sn	\hspace{5mm} &	3.010 	\hspace{5mm} &	3.137 	\hspace{5mm} &	3.353 	\hspace{5mm} &	3.315 	\hspace{5mm} &	1.300 	\hspace{5mm} &	3.339 	\hspace{5mm} &	3.196 	\hspace{5mm} &	1.195 	\\
Sr	\hspace{5mm} &	0.966 	\hspace{5mm} &	4.249 	\hspace{5mm} &	1.332 	\hspace{5mm} &	4.287 	\hspace{5mm} &	1.653 	\hspace{5mm} &	1.279 	\hspace{5mm} &	4.138 	\hspace{5mm} &	1.480 	\\
Ta	\hspace{5mm} &	6.963 	\hspace{5mm} &	2.729 	\hspace{5mm} &	8.102 	\hspace{5mm} &	2.811 	\hspace{5mm} &	1.266 	\hspace{5mm} &	8.089 	\hspace{5mm} &	2.712 	\hspace{5mm} &	1.121 	\\
Tc	\hspace{5mm} &	7.583 	\hspace{5mm} &	2.538 	\hspace{5mm} &	8.931 	\hspace{5mm} &	2.710 	\hspace{5mm} &	1.008 	\hspace{5mm} &	8.530 	\hspace{5mm} &	2.894 	\hspace{5mm} &	0.677 	\\
Ti	\hspace{5mm} &	4.450 	\hspace{5mm} &	2.674 	\hspace{5mm} &	5.469 	\hspace{5mm} &	2.933 	\hspace{5mm} &	1.030 	\hspace{5mm} &	5.536 	\hspace{5mm} &	2.795 	\hspace{5mm} &	0.930 	\\
Tl	\hspace{5mm} &	1.649 	\hspace{5mm} &	3.318 	\hspace{5mm} &	1.849 	\hspace{5mm} &	3.375 	\hspace{5mm} &	1.527 	\hspace{5mm} &	1.819 	\hspace{5mm} &	3.214 	\hspace{5mm} &	1.467 	\\
V 	\hspace{5mm} &	5.845 	\hspace{5mm} &	2.445 	\hspace{5mm} &	6.654 	\hspace{5mm} &	2.885 	\hspace{5mm} &	0.806 	\hspace{5mm} &	6.874 	\hspace{5mm} &	2.401 	\hspace{5mm} &	1.067 	\\
W 	\hspace{5mm} &	8.212 	\hspace{5mm} &	2.632 	\hspace{5mm} &	9.157 	\hspace{5mm} &	3.089 	\hspace{5mm} &	0.850 	\hspace{5mm} &	9.223 	\hspace{5mm} &	2.602 	\hspace{5mm} &	1.117 	\\
Y 	\hspace{5mm} &	2.779 	\hspace{5mm} &	3.424 	\hspace{5mm} &	3.672 	\hspace{5mm} &	3.567 	\hspace{5mm} &	1.346 	\hspace{5mm} &	3.590 	\hspace{5mm} &	3.493 	\hspace{5mm} &	1.168 	\\
Zn	\hspace{5mm} &	0.862 	\hspace{5mm} &	2.538 	\hspace{5mm} &	1.039 	\hspace{5mm} &	2.569 	\hspace{5mm} &	1.308 	\hspace{5mm} &	0.817 	\hspace{5mm} &	2.537 	\hspace{5mm} &	1.194 	\\
Zr	\hspace{5mm} &	5.096 	\hspace{5mm} &	2.918 	\hspace{5mm} &	6.059 	\hspace{5mm} &	3.203 	\hspace{5mm} &	1.166 	\hspace{5mm} &	6.112 	\hspace{5mm} &	3.016 	\hspace{5mm} &	1.077 	\\

\hline
\end{tabular}
}
\label{table1}
\end{center}
\end{table*}

\section{Structural parameters}
\label{sec:list}

The cohesive energy of the structure $j$ is defined as $E_{{\rm c}, j} = E_{\rm atom} - E_{j}$, where $E_{\rm atom}$ is the total energy of an atom in free space and $E_{j}$ is the total energy per atom of the structure $j$. During the geometry optimization for the structure $j$, the total energy and forces are converged within $10^{-5}$ Ry and $10^{-4}$ a.u., respectively. The value of $E_{\rm atom}$ for all atoms is obtained from atom-in-a-box calculations (within non-spin-polarized approximation) of a single atom in a unit cell with a volume of 15$\times$15$\times$15 \AA$^3$. Table \ref{table1} lists $E_{{\rm c}, j}$, the lattice constant $a_j$, and the buckling height $\delta_j$, with the subscripts $j=1$ for the HX, $j=2$ for the bHC or 2HC, and $j=3$ for the bSQ or 2SQ structures, where the data of $\delta_1$ that is equal to zero is omitted. 

For 3SQ V, 3SQ Nb, 3SQ Ta, and 3HX In, the cohesive energy $E_{{\rm c},4}$ (eV), the lattice constant $a_4$ (\AA), and the interlayer distance $2\delta_4$ (\AA) are $(E_{{\rm c},4},a_4,2\delta_4)=(6.947, 2.886, 1.404)$, (8.209, 3.256, 1.471), (8.195, 3.258, 1.469), and (2.223, 3.315, 2.768), respectively. 

For AlCu alloys, the optimized lattice constants ($a, c$) are $(2.787, 4.077)$ \AA \ for B$_h$ structure ($c/a=1.463$) and $(2.900, 3.202)$ \AA \ for L1$_0$ structure ($c/a=1.104$). For B$_h$ CuZn, the lattice parameters ($a, c$) are $(2.676, 4.211)$ \AA, i.e., $c/a=1.574$. 


\section{Phonon band structures}
\label{sec:results}
Figures \ref{fig_Li}-\ref{fig_Pb} show the phonon band structures from Li to Pb for (a) HX, (b) bHC, and (c) bSQ structures. The phonon band structure and the trend of the dynamical stability of HX, bHC, and bSQ are similar in the same group in the periodic table. Below, the results for non-spin-polarized calculations are shown. The phonon band structures of bHC Co (spin-polarized), bHC Ni (spin-polarized), trilayer V, Nb, Ta, Ga, In, and Tl will be shown at the end of SM. The stability properties of elemental metals are summarized in the periodic table shown in Fig.~\ref{fig_S1}. 

\subsection*{Group 1: Li, Na, K, Rb, and Cs (alkali metals)}
\begin{figure}[H] 
\center
\includegraphics[scale=0.30]{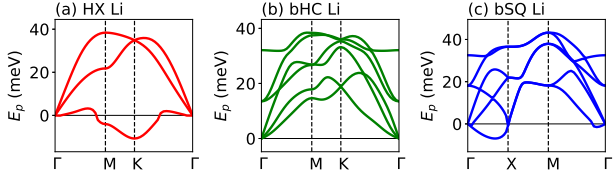}
\caption{The phonon band structure of 2D Li for (a) HX, (b) bHC, and (c) bSQ structures. }
\label{fig_Li} 
\end{figure}

\begin{figure}[H]
\center
\includegraphics[scale=0.30]{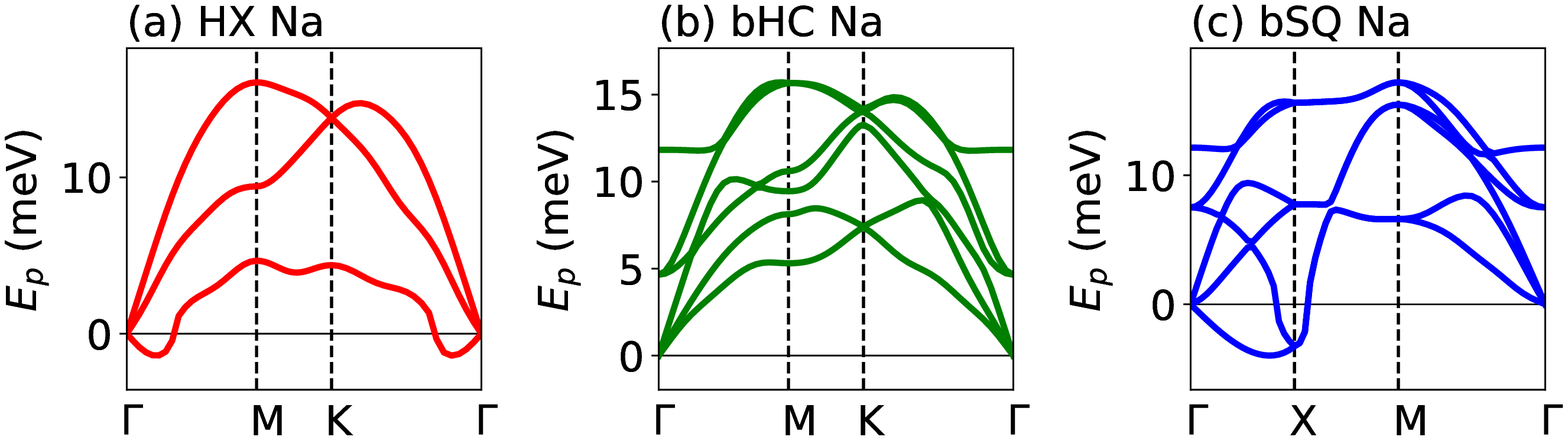}
\caption{Same as Fig.~\ref{fig_Li} but for Na.}\label{fig_Na} 
\end{figure}

\begin{figure}[H]
\center
\includegraphics[scale=0.30]{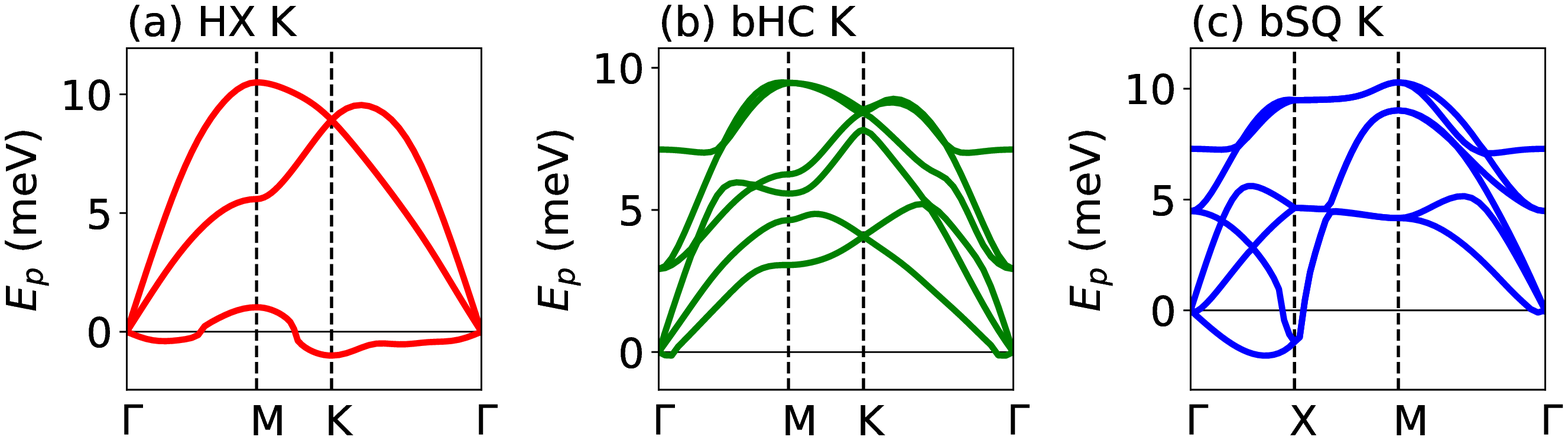}
\caption{Same as Fig.~\ref{fig_Li} but for K.}\label{fig_K} 
\end{figure}

\begin{figure}[H]
\center
\includegraphics[scale=0.30]{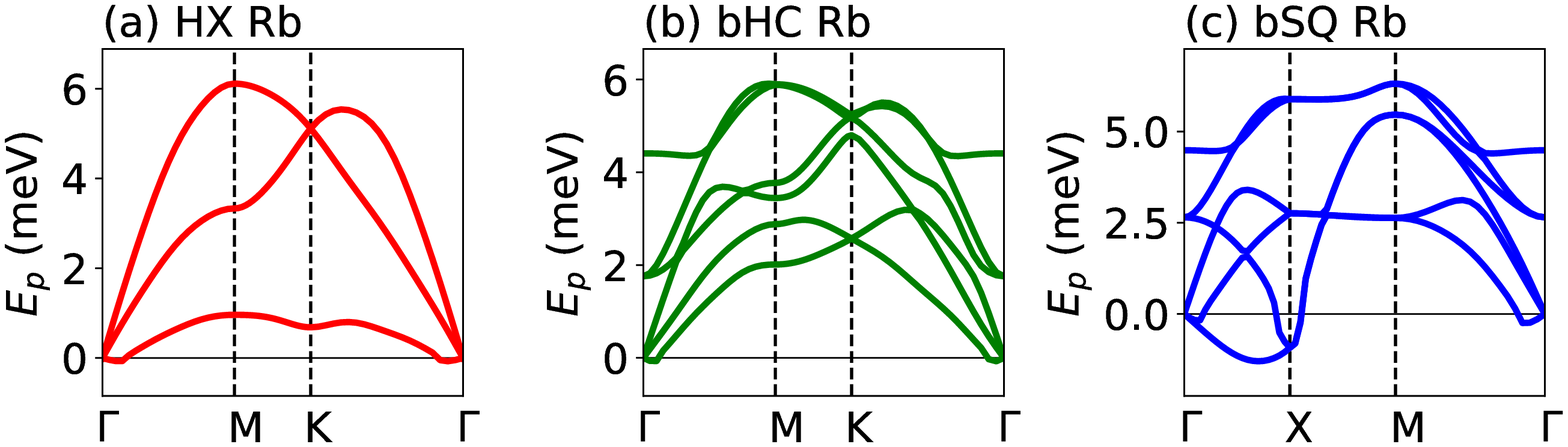}
\caption{Same as Fig.~\ref{fig_Li} but for Rb.}\label{fig_Rb} 
\end{figure}

\begin{figure}[H]
\center
\includegraphics[scale=0.30]{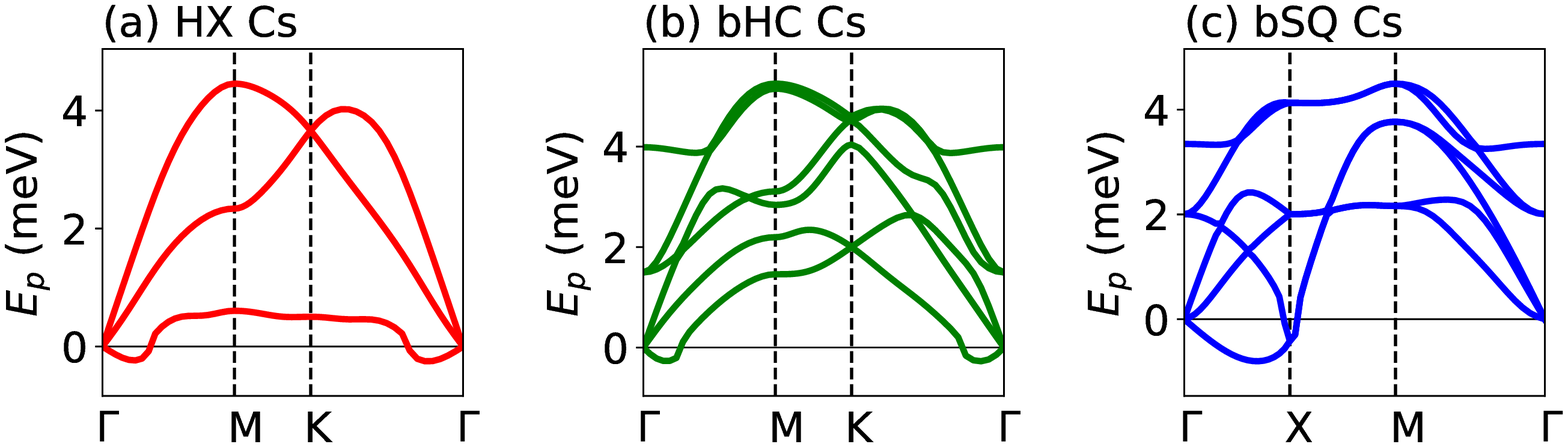}
\caption{Same as Fig.~\ref{fig_Li} but for Cs.}\label{fig_Cs} 
\end{figure}

Figures \ref{fig_Li}, \ref{fig_Na}, \ref{fig_K}, \ref{fig_Rb}, and \ref{fig_Cs} show the phonon band structures of Li, Na, K, Rb, and Cs, respectively. The HX Rb is dynamically stable, while HX Li, Na, K, and Cs are unstable, since the imaginary frequencies in the ZA branch are observed around $\Gamma$. When the bHC phase is assumed, the ZA phonon modes are stabilized. This yields the dynamical stability of 2D Li, Na, and K, while the bHC Cs is still unstable. The alkali metals in the bSQ structure are all unstable.  

\subsection*{Group 2: Be, Mg, Ca, Sr, and Ba (alkali earth metals)}
\begin{figure}[H]
\center
\includegraphics[scale=0.30]{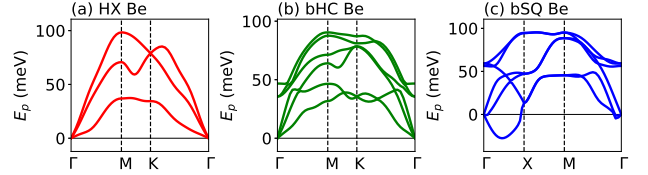}
\caption{Same as Fig.~\ref{fig_Li} but for Be.}\label{fig_Be} 
\end{figure}

\begin{figure}[H]
\center
\includegraphics[scale=0.30]{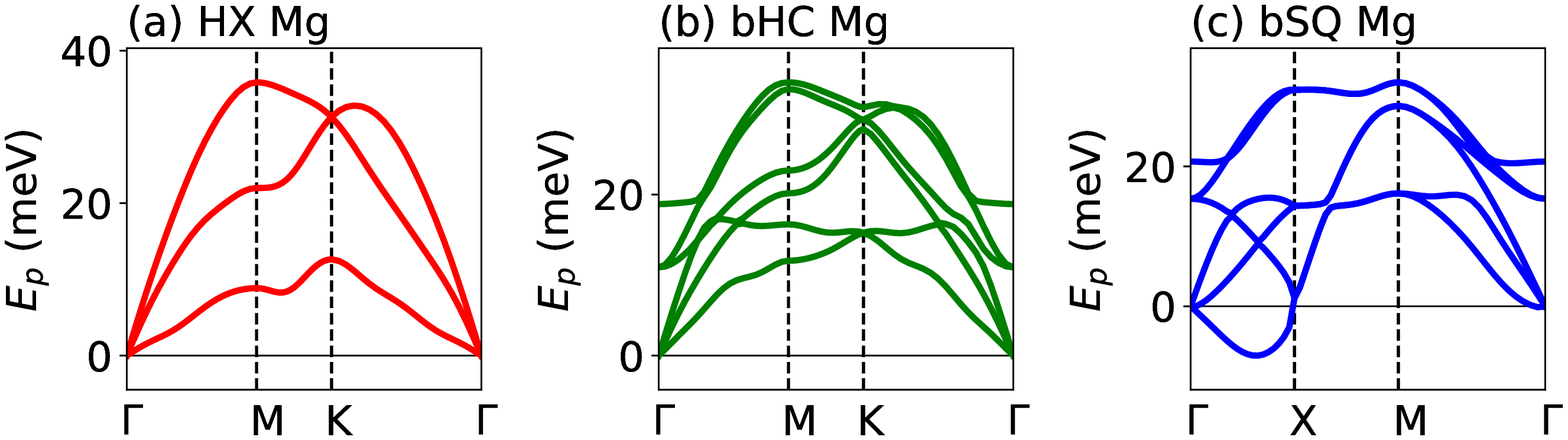}
\caption{Same as Fig.~\ref{fig_Li} but for Mg.}\label{fig_Mg} 
\end{figure}

\begin{figure}[H]
\center
\includegraphics[scale=0.30]{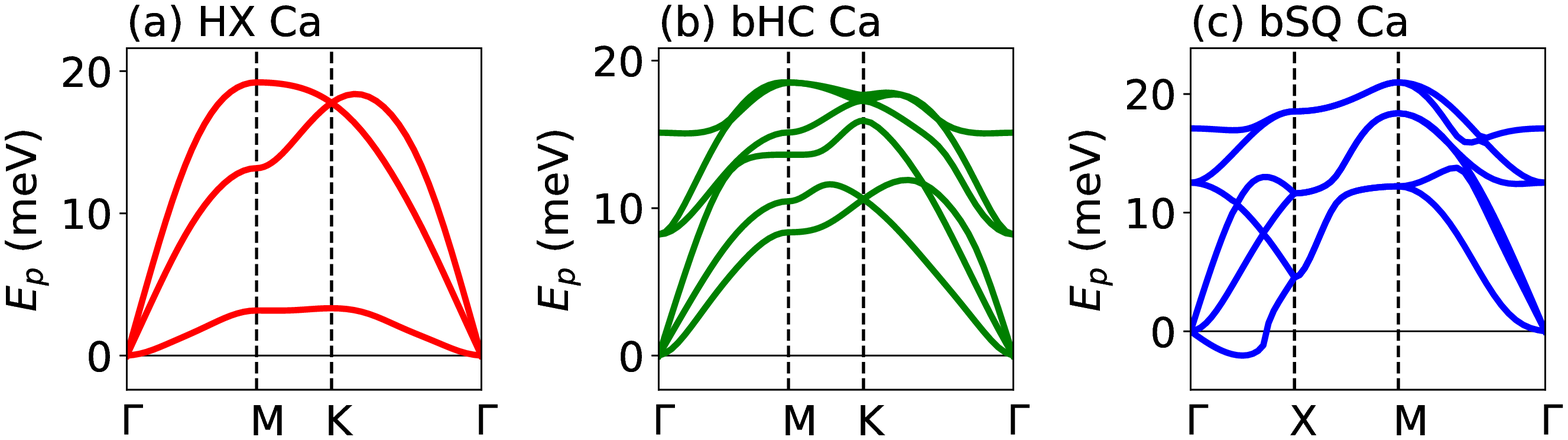}
\caption{Same as Fig.~\ref{fig_Li} but for Ca.}\label{fig_Ca} 
\end{figure}

\begin{figure}[H]
\center
\includegraphics[scale=0.30]{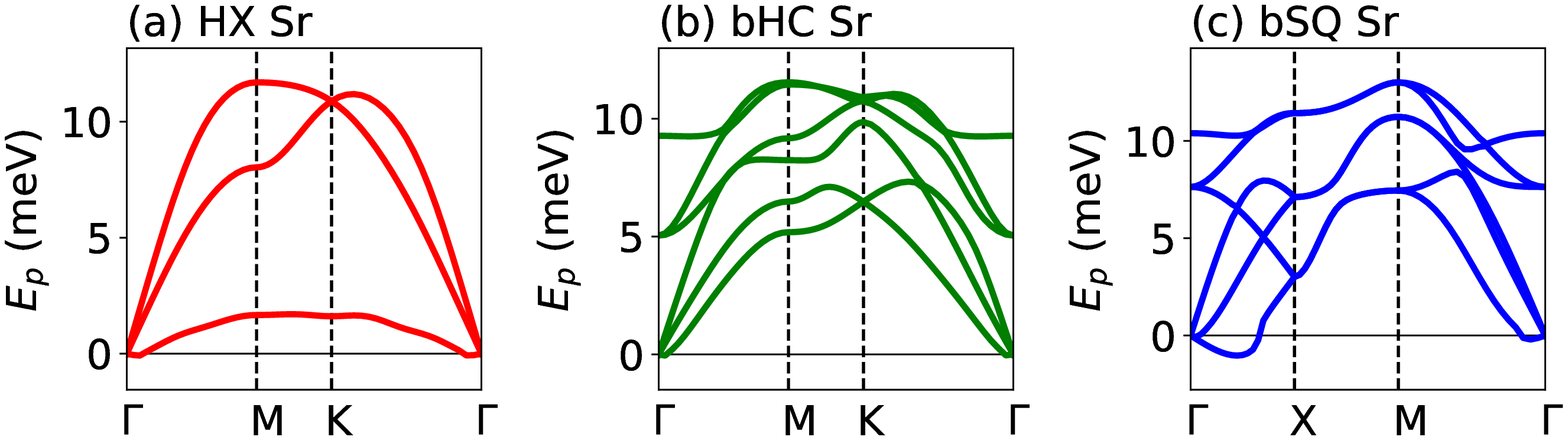}
\caption{Same as Fig.~\ref{fig_Li} but for Sr.}\label{fig_Sr} 
\end{figure}

\begin{figure}[H]
\center
\includegraphics[scale=0.30]{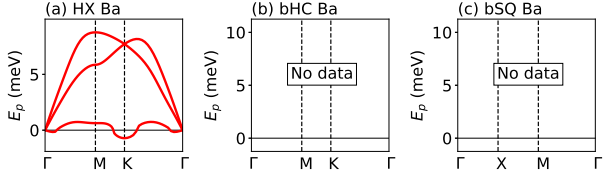}
\caption{Same as Fig.~\ref{fig_Li} but for Ba.}\label{fig_Ba} 
\end{figure}

Figures \ref{fig_Be}, \ref{fig_Mg}, \ref{fig_Ca}, \ref{fig_Sr}, and \ref{fig_Ba} show the phonon band structures of Be, Mg, Ca, Sr, and Ba, respectively. The HX and bHC structures are dynamically stable except Ba. As the ion mass increases in the HX structure, the phonon softening behavior is observed in ZA branch. The maximum phonon energy is higher than that of the alkali metal in the same period because as listed in Table \ref{table1} the values of $E_{{\rm c}, j}$ and $a_j$ in the alkali earth metals are larger and smaller than those in the alkali metals, respectively. The bSQ structure is dynamically unstable in alkali earth metals. The band structures of bHC and bSQ Ba are omitted due to no convergence in the scf calculations. 


\subsection*{Group 3: Sc, Y, and Lu}
\begin{figure}[H]
\center
\includegraphics[scale=0.30]{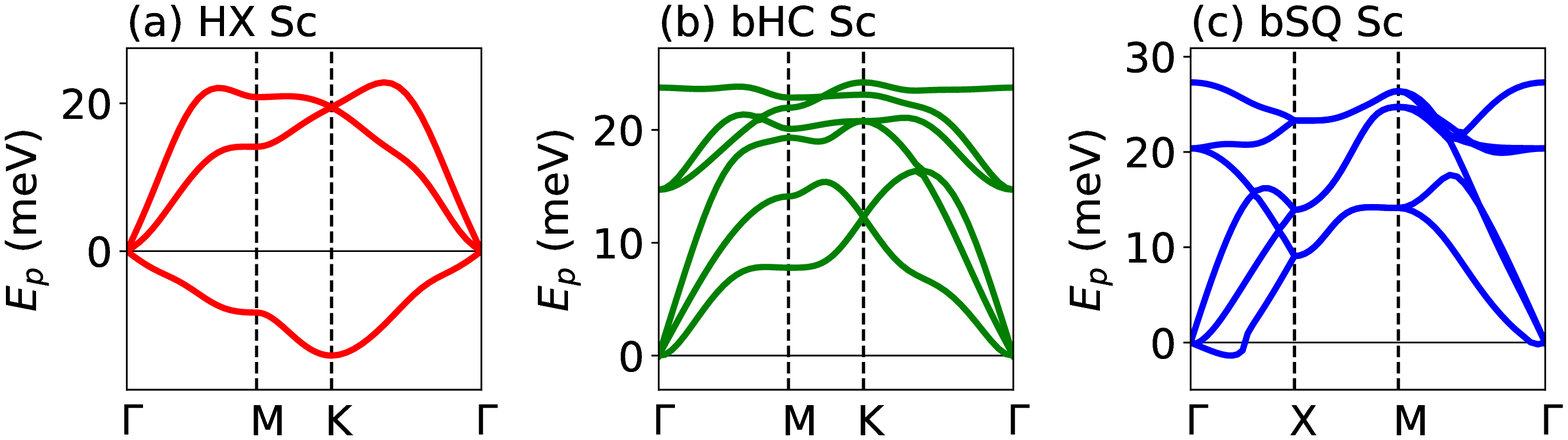}
\caption{\label{fig_Sc} Same as Fig.~\ref{fig_Li} but for Sc.}
\end{figure}

\begin{figure}[H]
\center
\includegraphics[scale=0.30]{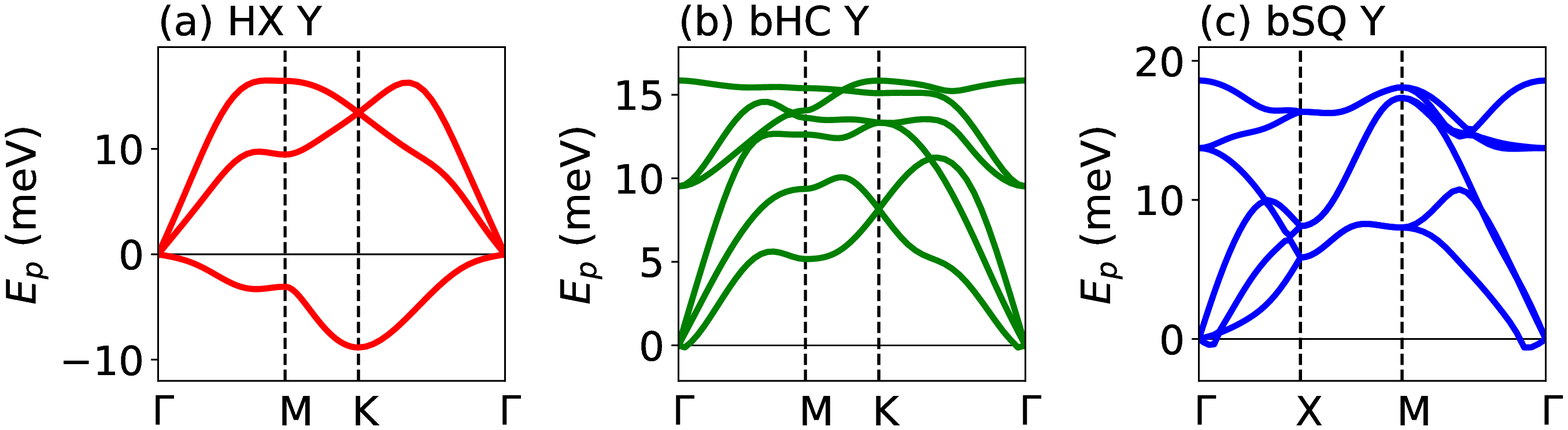}
\caption{\label{fig_Y} Same as Fig.~\ref{fig_Li} but for Y.}
\end{figure}

\begin{figure}[H]
\center
\includegraphics[scale=0.30]{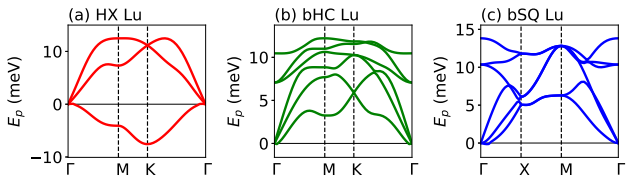}
\caption{\label{fig_Lu} Same as Fig.~\ref{fig_Li} but for Lu.}
\end{figure}

Figures \ref{fig_Sc}, \ref{fig_Y}, and \ref{fig_Lu} show the phonon band structures of Sc, Y, and Lu, respectively. The bHC and bSQ structures are dynamically stable. Since the bHC is energetically more stable than the bSQ structure as listed in Table \ref{table1}, the bSQ is regarded as a metastable state. 

\subsection*{Group 4: Ti, Zr, and Hf}

\begin{figure}[H]
\center
\includegraphics[scale=0.30]{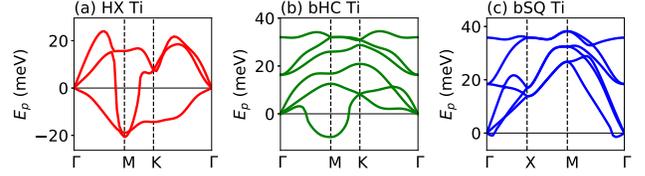}
\caption{\label{fig_Ti} Same as Fig.~\ref{fig_Li} but for Ti.}
\end{figure}

\begin{figure}[H]
\center
\includegraphics[scale=0.30]{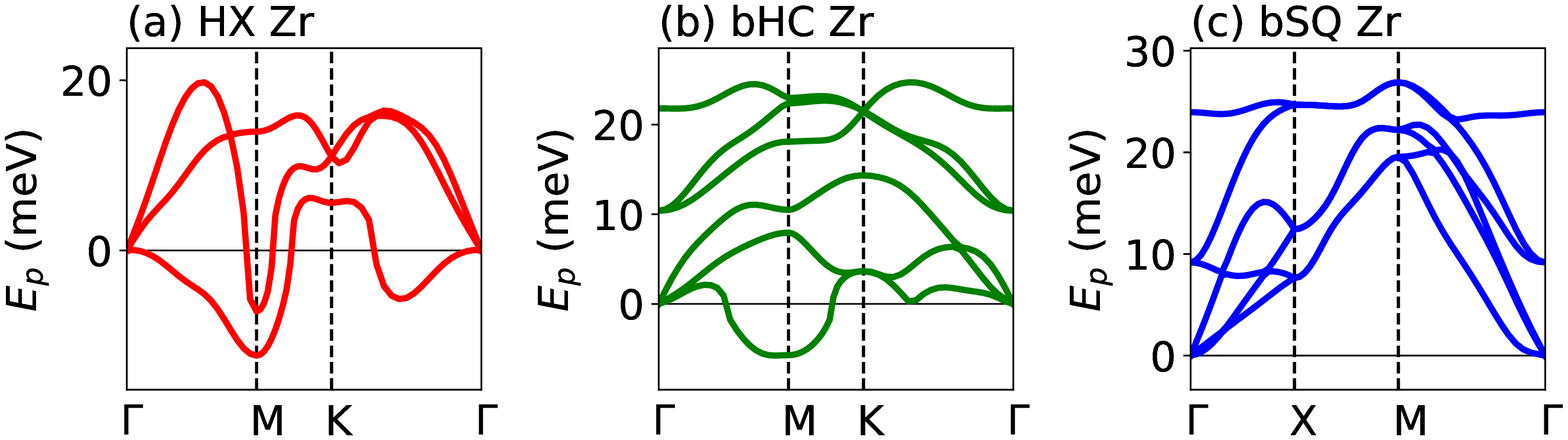}
\caption{\label{fig_Zr} Same as Fig.~\ref{fig_Li} but for Zr.}
\end{figure}

\begin{figure}[H]
\center
\includegraphics[scale=0.30]{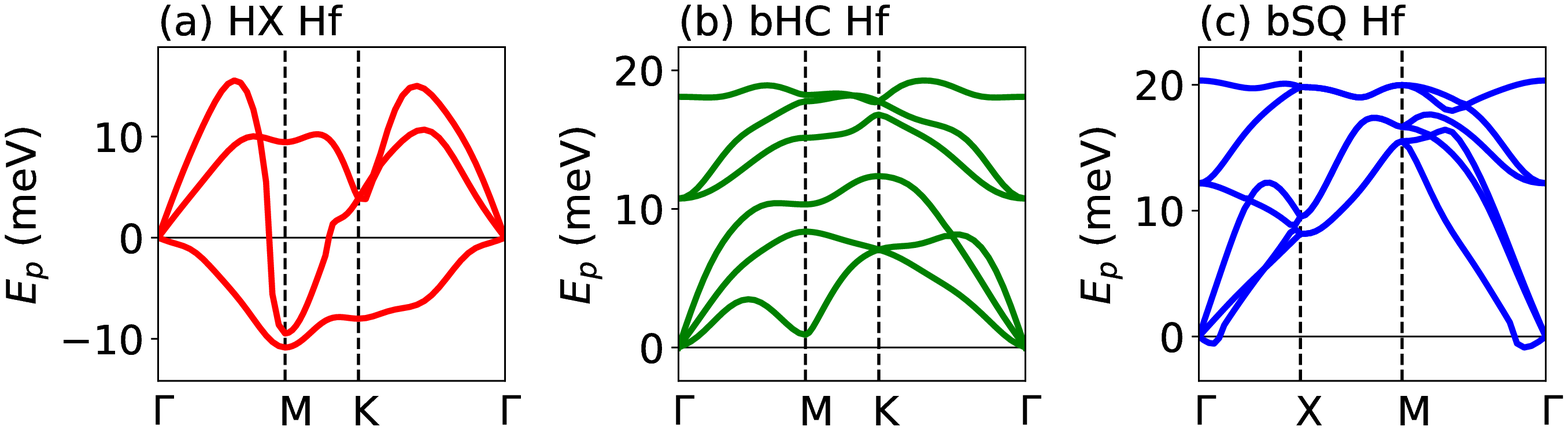}
\caption{\label{fig_Hf} Same as Fig.~\ref{fig_Li} but for Hf.}
\end{figure}

Figures \ref{fig_Ti}, \ref{fig_Zr}, and \ref{fig_Hf} show the phonon band structures of Ti, Zr, and Hf, respectively. The bSQ structure is dynamically stable for these elements. For Hf, the bHC is energetically more stable than the bSQ structure by 9 meV per unit cell, which gives the dynamical stability in the bHC Hf. 

\subsection*{Group 5: V, Nb, and Ta}

\begin{figure}[H]
\center
\includegraphics[scale=0.30]{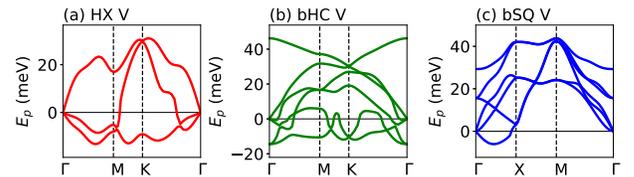}
\caption{\label{fig_V} Same as Fig.~\ref{fig_Li} but for V.}
\end{figure}

\begin{figure}[H]
\center
\includegraphics[scale=0.30]{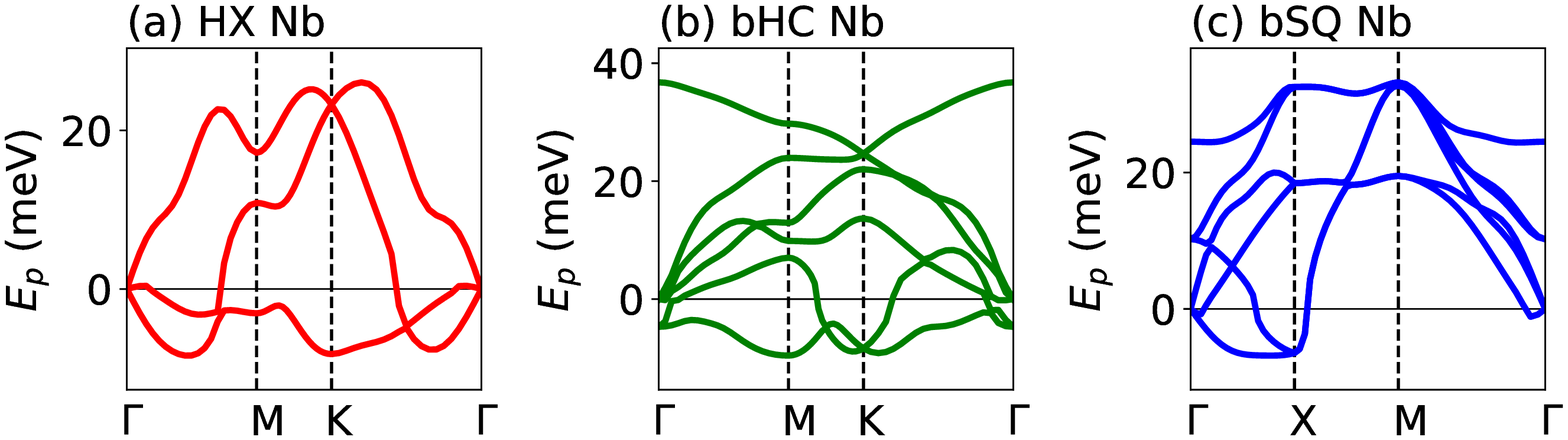}
\caption{\label{fig_Nb} Same as Fig.~\ref{fig_Li} but for Nb.}
\end{figure}

\begin{figure}[H]
\center
\includegraphics[scale=0.30]{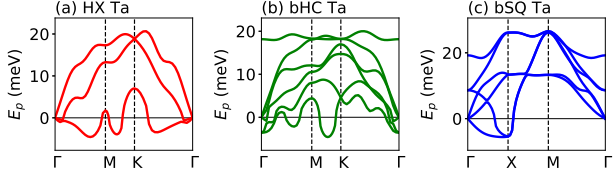}
\caption{\label{fig_Ta} Same as Fig.~\ref{fig_Li} but for Ta.}
\end{figure}

Figures \ref{fig_V}, \ref{fig_Nb}, and \ref{fig_Ta} show the phonon band structures of V, Nb, and Ta, respectively. No stable structures are obtained. As shown later, the 3SQ structure is dynamically stable.

\subsection*{Group 6: Cr, Mo, and W}

\begin{figure}[H]
\center
\includegraphics[scale=0.30]{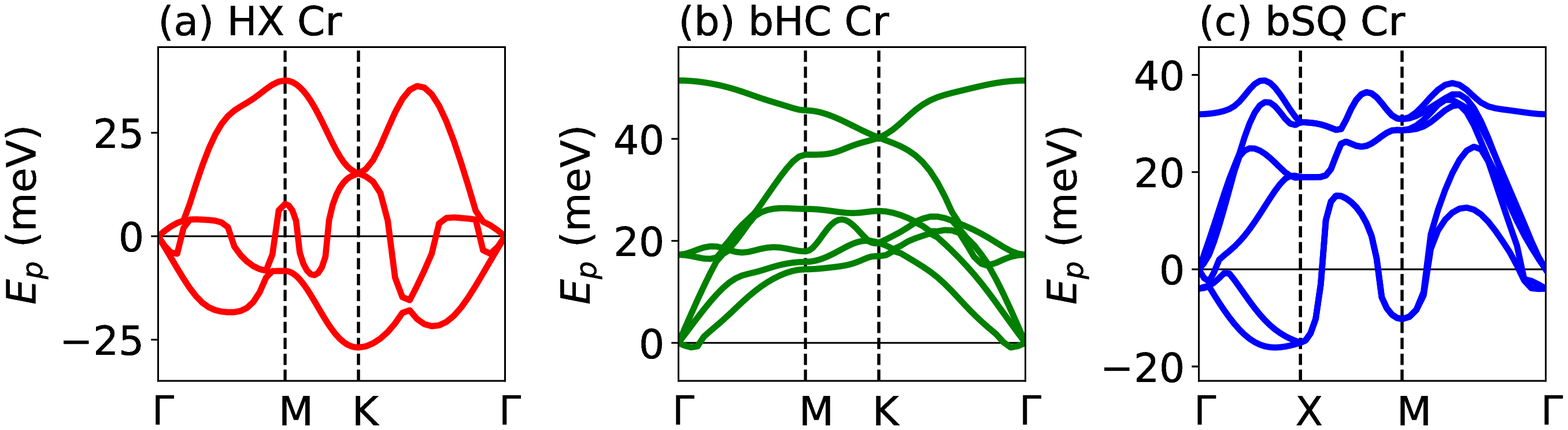}
\caption{\label{fig_Cr} Same as Fig.~\ref{fig_Li} but for Cr.}
\end{figure}

\begin{figure}[H]
\center
\includegraphics[scale=0.30]{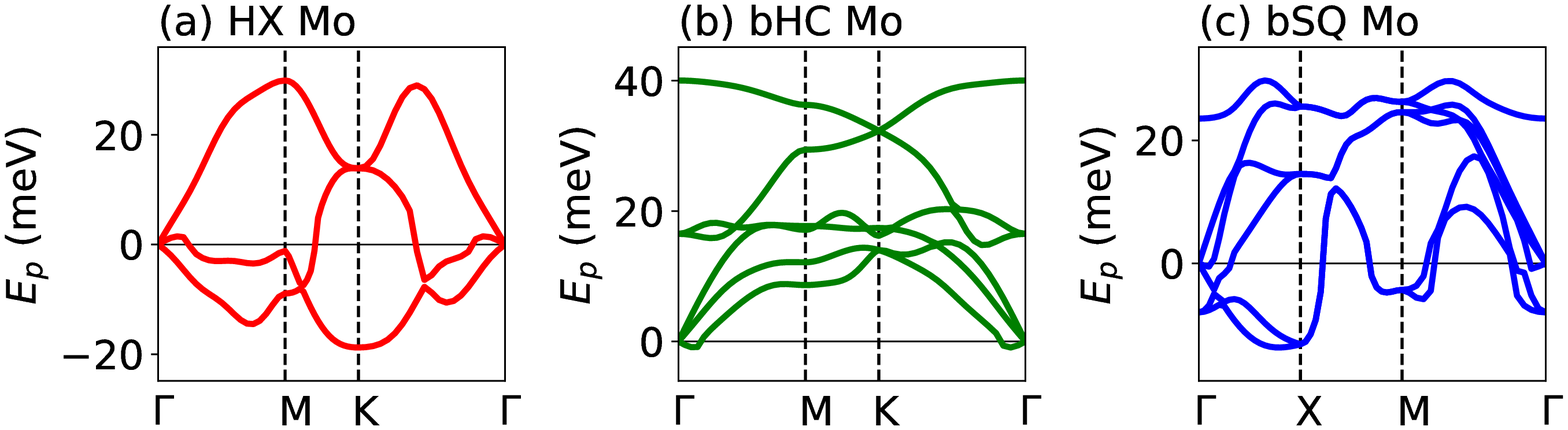}
\caption{\label{fig_Mo} Same as Fig.~\ref{fig_Li} but for Mo.}
\end{figure}

\begin{figure}[H]
\center
\includegraphics[scale=0.30]{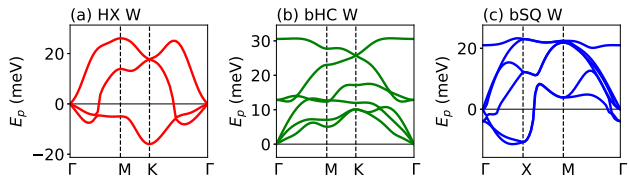}
\caption{\label{fig_W} Same as Fig.~\ref{fig_Li} but for W.}
\end{figure}

Figures \ref{fig_Cr}, \ref{fig_Mo}, and \ref{fig_W} show the phonon band structures of Cr, Mo, and W, respectively. These elements in the bHC structure are dynamically stable. It should be noted that for Mo and W, we have found other bHC structure different from the bHC structure listed in Table \ref{table1}. Although the former has a larger $E_{{\rm c},2}$, smaller $a_2$, and higher buckling height $\delta_2$, such a structure is dynamically unstable.  

\subsection*{Group 7: Mn, Tc, and Re}

\begin{figure}[H]
\center
\includegraphics[scale=0.30]{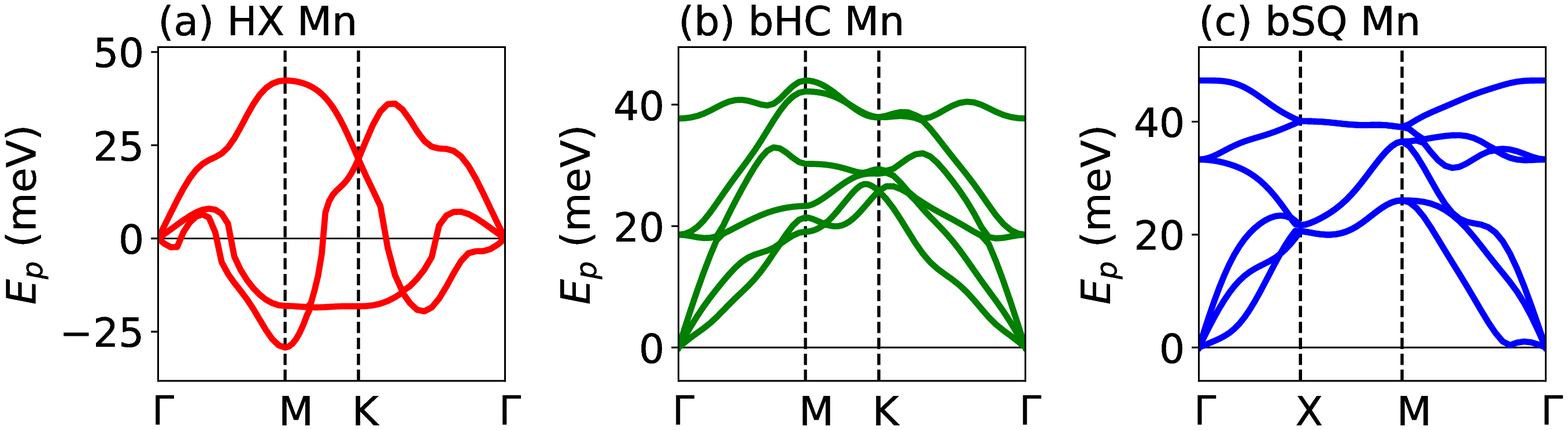}
\caption{\label{fig_Mn} Same as Fig.~\ref{fig_Li} but for Mn.}
\end{figure}

\begin{figure}[H]
\center
\includegraphics[scale=0.30]{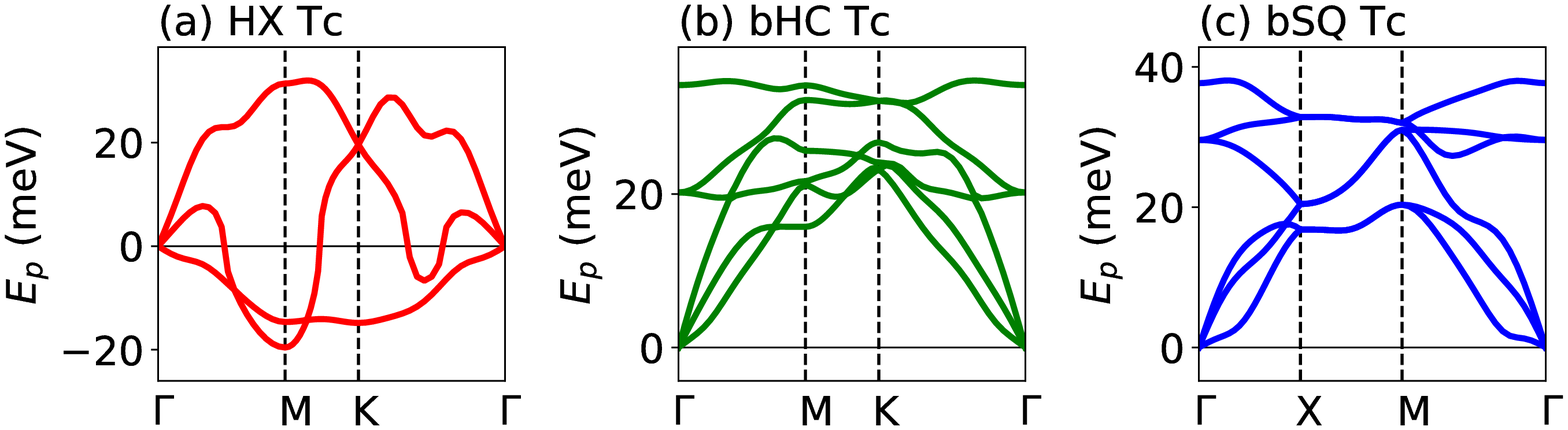}
\caption{\label{fig_Tc} Same as Fig.~\ref{fig_Li} but for Tc.}
\end{figure}

\begin{figure}[H]
\center
\includegraphics[scale=0.30]{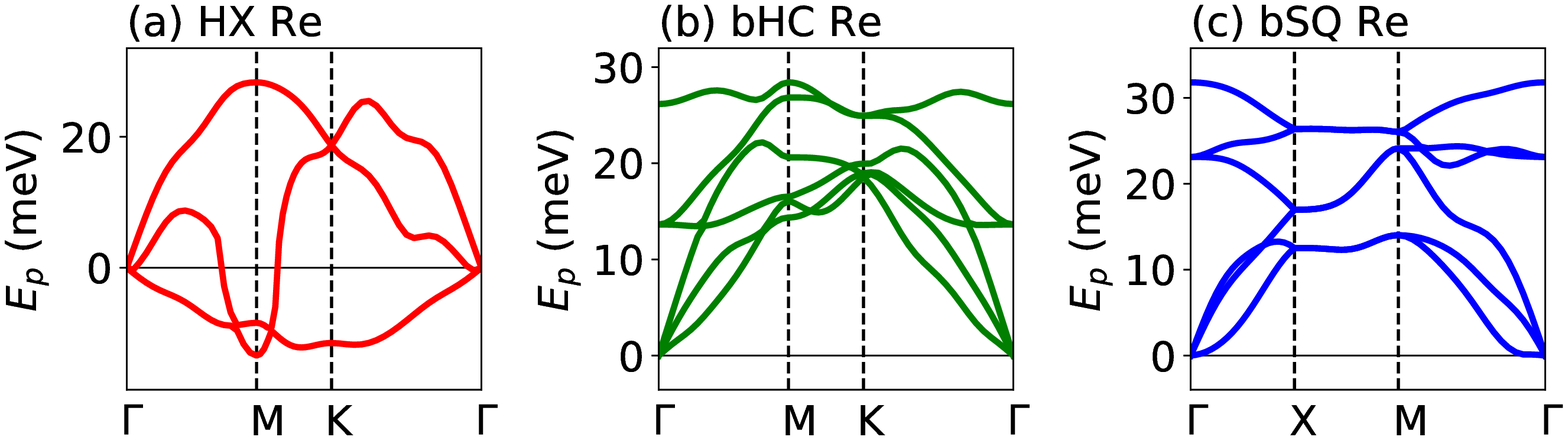}
\caption{\label{fig_Re} Same as Fig.~\ref{fig_Li} but for Re.}
\end{figure}

Figures \ref{fig_Mn}, \ref{fig_Tc}, and \ref{fig_Re} show the phonon band structures of Mn, Tc, and Re, respectively. The bHC and bSQ structures are dynamically stable in these elements, whereas the bSQ structure is metastable. 

\subsection*{Group 8: Fe, Ru, and Os}

\begin{figure}[H]
\center
\includegraphics[scale=0.30]{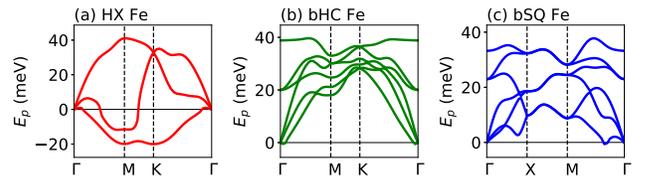}
\caption{\label{fig_Fe} Same as Fig.~\ref{fig_Li} but for Fe.}
\end{figure}

\begin{figure}[H]
\center
\includegraphics[scale=0.30]{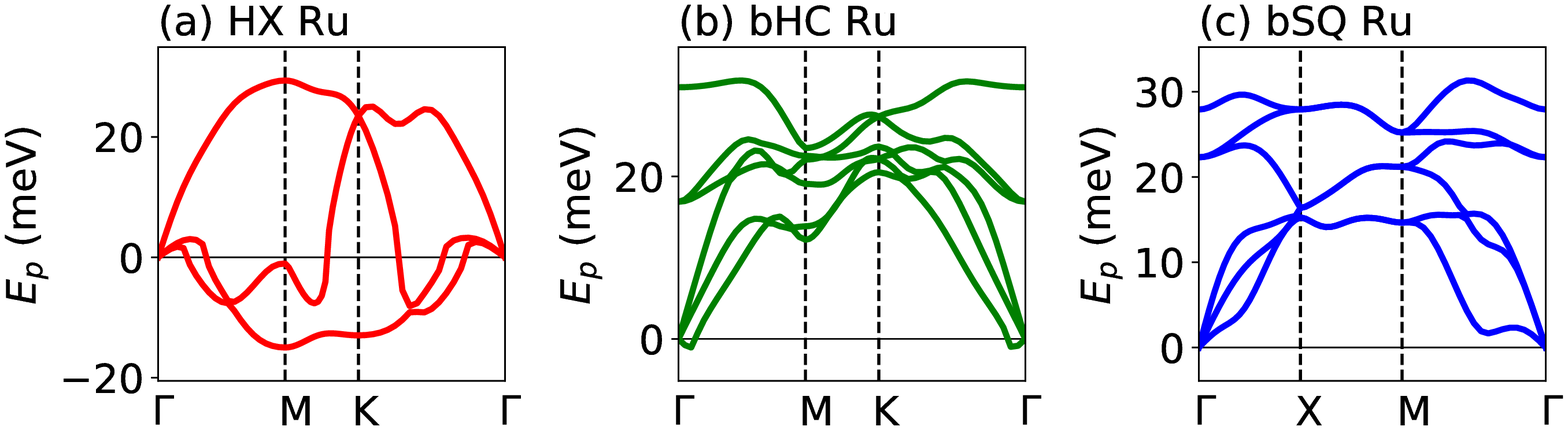}
\caption{\label{fig_Ru} Same as Fig.~\ref{fig_Li} but for Ru.}
\end{figure}

\begin{figure}[H]
\center
\includegraphics[scale=0.30]{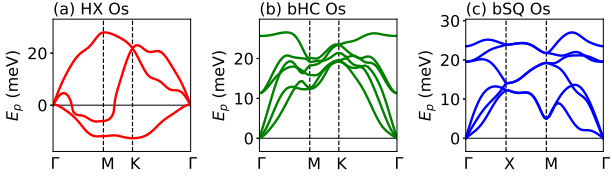}
\caption{\label{fig_Os} Same as Fig.~\ref{fig_Li} but for Os.}
\end{figure}

Figures \ref{fig_Fe}, \ref{fig_Ru}, and \ref{fig_Os} show the phonon band structures of Fe, Ru, and Os, respectively. These have the bHC and bSQ as dynamically stable structures, while the bSQ is a metastable structure. 

\subsection*{Group 9: Co, Rh, and Ir}

\begin{figure}[H]
\center
\includegraphics[scale=0.30]{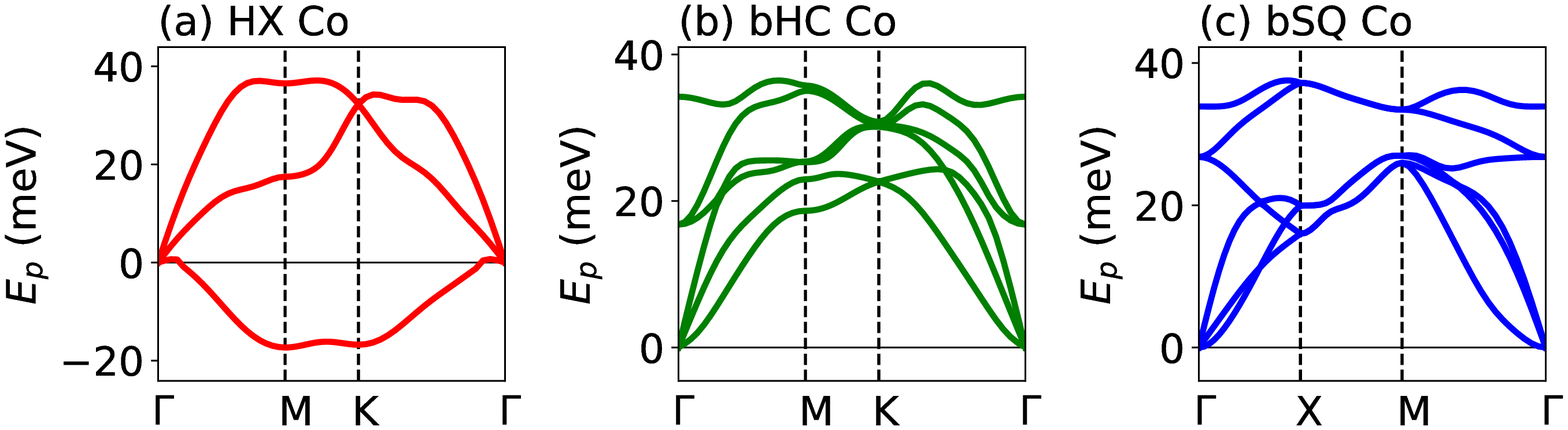}
\caption{\label{fig_Co} Same as Fig.~\ref{fig_Li} but for Co.}
\end{figure}

\begin{figure}[H]
\center
\includegraphics[scale=0.30]{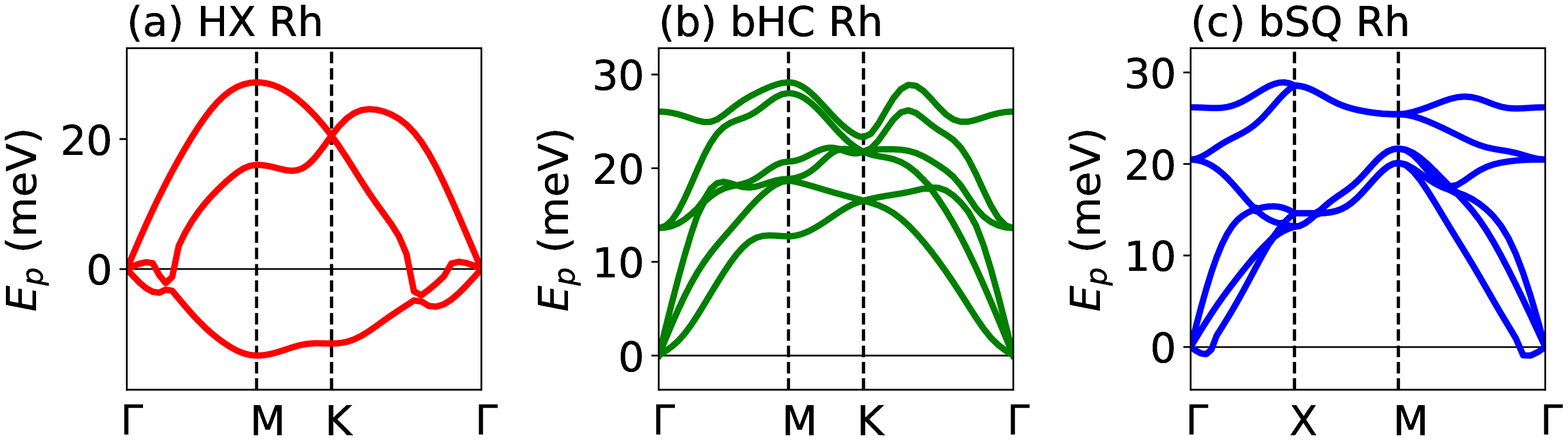}
\caption{\label{fig_Rh} Same as Fig.~\ref{fig_Li} but for Rh.}
\end{figure}

\begin{figure}[H]
\center
\includegraphics[scale=0.30]{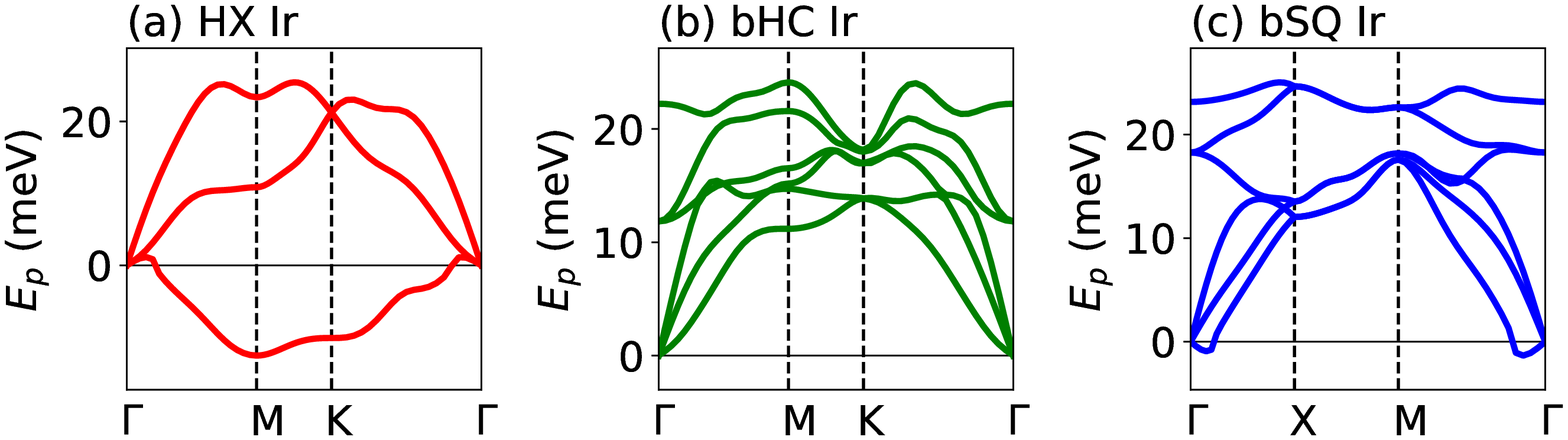}
\caption{\label{fig_Ir} Same as Fig.~\ref{fig_Li} but for Ir.}
\end{figure}

Figures \ref{fig_Co}, \ref{fig_Rh}, and \ref{fig_Ir} show the phonon band structures of Co, Rh, and Ir, respectively. These elements in the bHC and bSQ structures are dynamically stable, while the bSQ structure is metastable.  

\subsection*{Group 10: Ni, Pd, and Pt}

\begin{figure}[H]
\center
\includegraphics[scale=0.30]{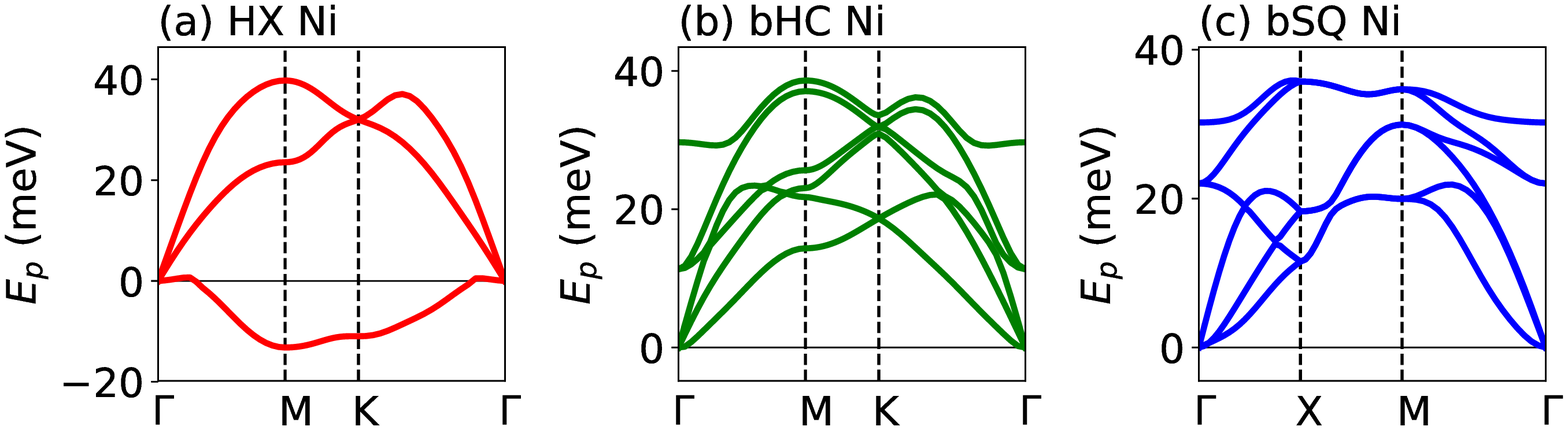}
\caption{\label{fig_Ni} Same as Fig.~\ref{fig_Li} but for Ni.}
\end{figure}

\begin{figure}[H]
\center
\includegraphics[scale=0.30]{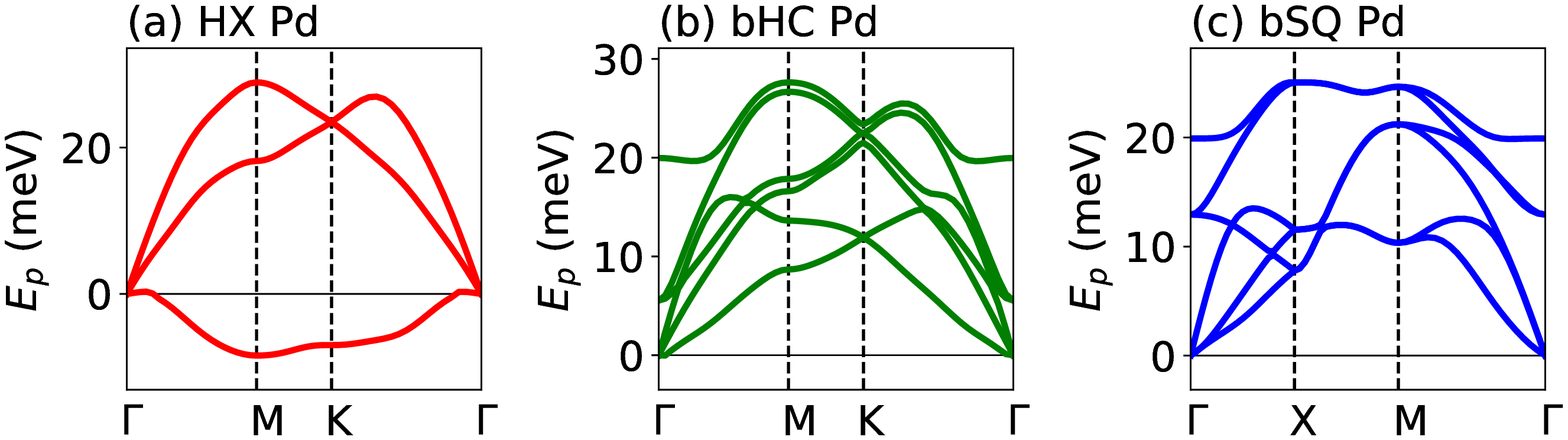}
\caption{\label{fig_Pd} Same as Fig.~\ref{fig_Li} but for Pd.}
\end{figure}

\begin{figure}[H]
\center
\includegraphics[scale=0.30]{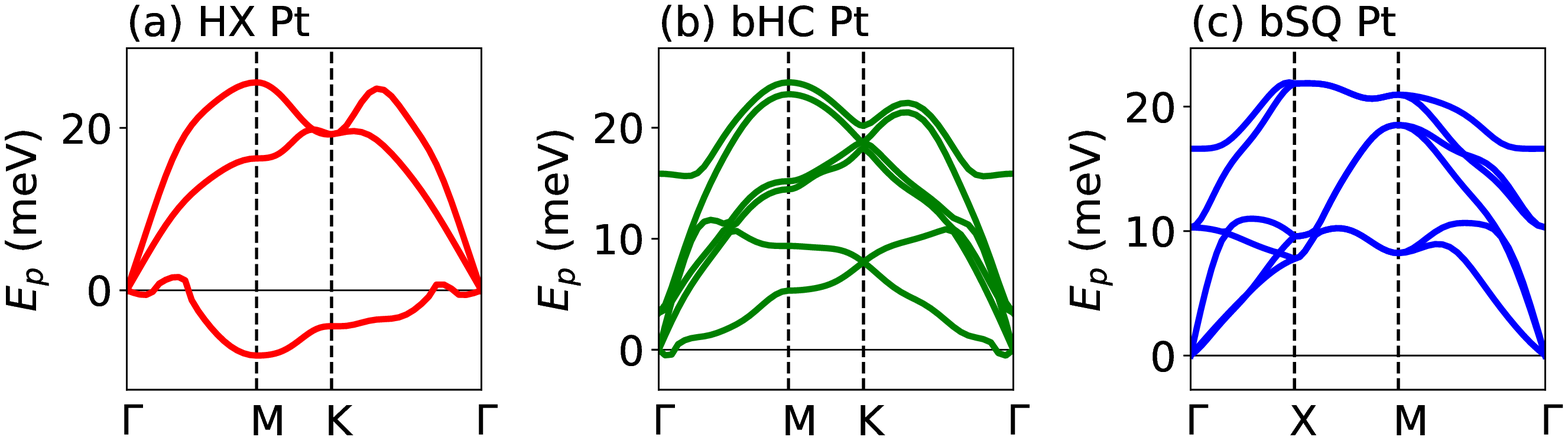}
\caption{Same as Fig.~\ref{fig_Li} but for Pt.}\label{fig_Pt} 
\end{figure}

Figures \ref{fig_Ni}, \ref{fig_Pd}, and \ref{fig_Pt} show the phonon band structures of Ni, Pd, and Pt, respectively. These elements in the bHC and bSQ structures are dynamically stable, while the bSQ is a metastable structure.  

\subsection*{Group 11: Cu, Ag, and Au (nobel metals)}

\begin{figure}[H]
\center
\includegraphics[scale=0.30]{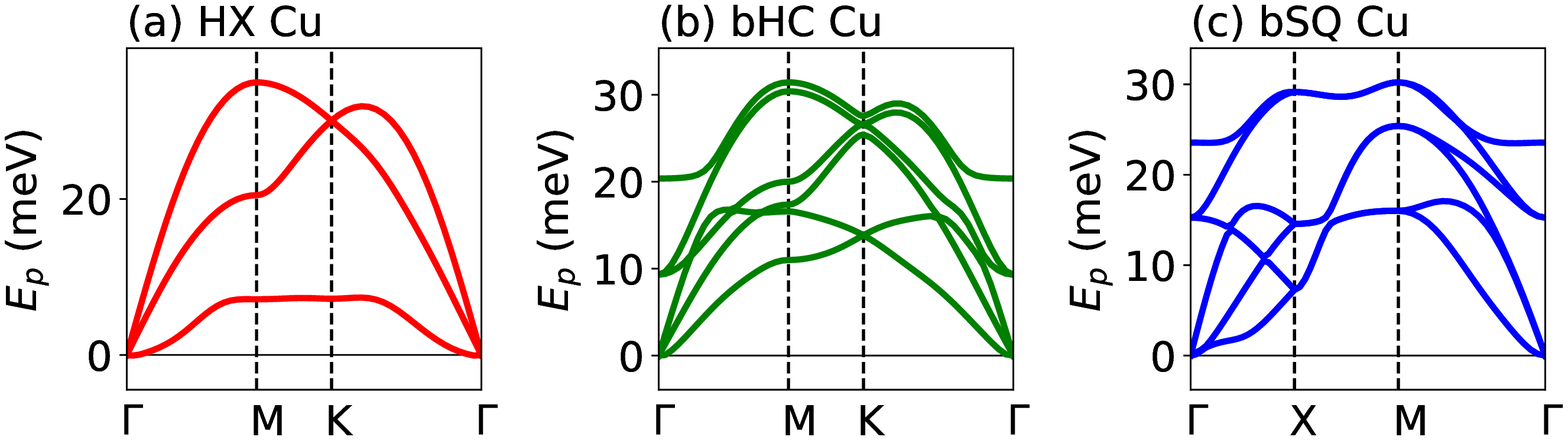}
\caption{Same as Fig.~\ref{fig_Li} but for Cu.}\label{fig_Cu} 
\end{figure}

\begin{figure}[H]
\center
\includegraphics[scale=0.30]{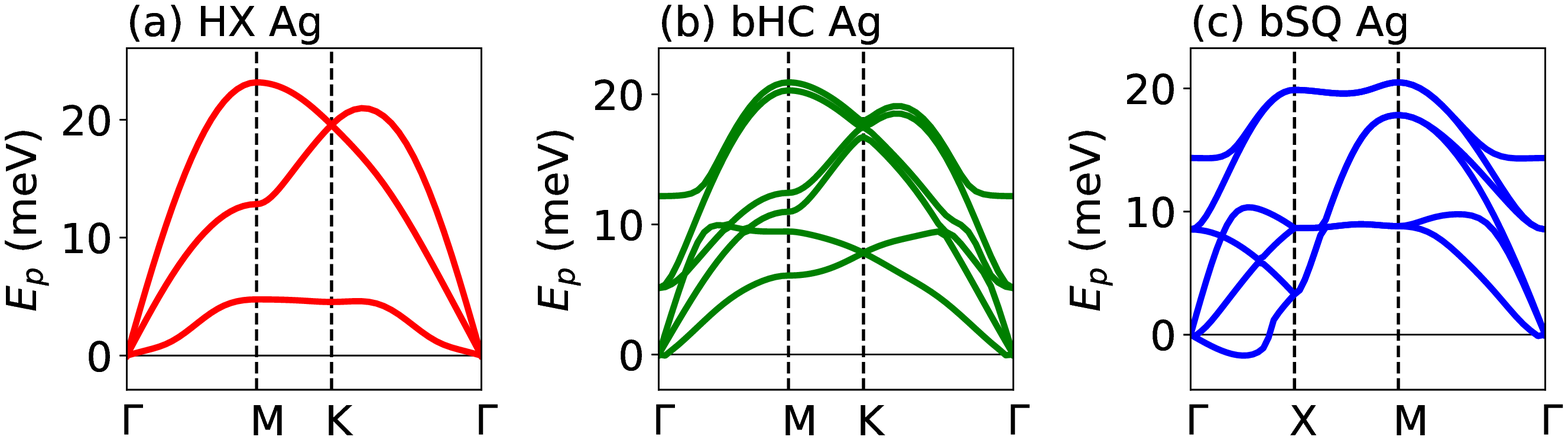}
\caption{Same as Fig.~\ref{fig_Li} but for Ag.}\label{fig_Ag} 
\end{figure}

\begin{figure}[H]
\center
\includegraphics[scale=0.30]{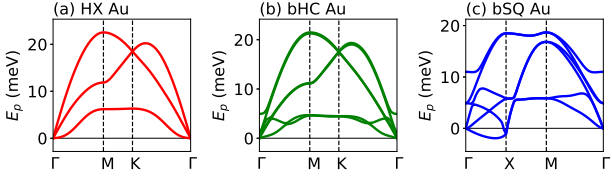}
\caption{Same as Fig.~\ref{fig_Li} but for Au.}\label{fig_Au}
\end{figure}

Figures \ref{fig_Cu}, \ref{fig_Ag} (the same as in Fig.~3 in the main text), and \ref{fig_Au} show the phonon band structure of Cu, Ag, and Au, respectively. The HX and bHC structures are dynamically stable. The former result (the stability of HX) is consistent with the previous calculations [11,12,13]. Cu is dynamically stable in the form of bSQ structure and is the only element with no imaginary frequencies observed.   

\subsection*{Group 12: Zn, Cd, and Hg}

\begin{figure}[H]
\center
\includegraphics[scale=0.30]{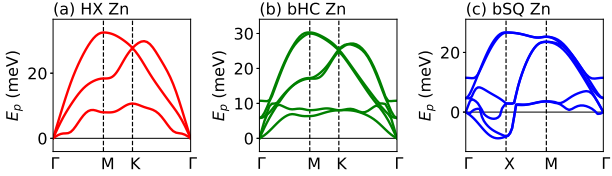}
\caption{\label{fig_Zn} Same as Fig.~\ref{fig_Li} but for Zn.}
\end{figure}

\begin{figure}[H]
\center
\includegraphics[scale=0.30]{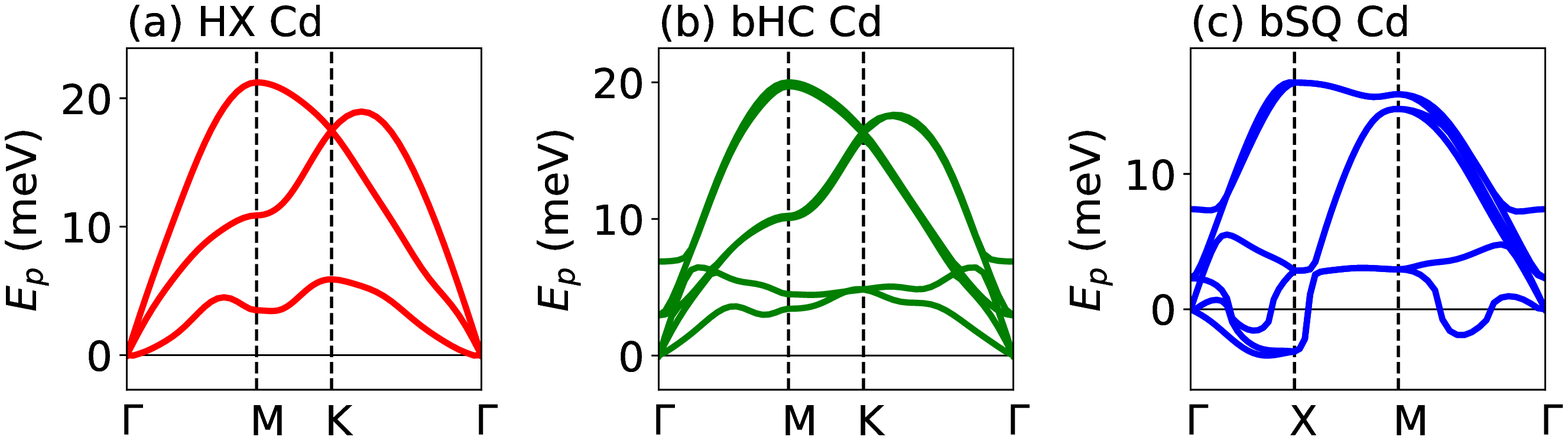}
\caption{\label{fig_Cd} Same as Fig.~\ref{fig_Li} but for Cd.}
\end{figure}

\begin{figure}[H]
\center
\includegraphics[scale=0.30]{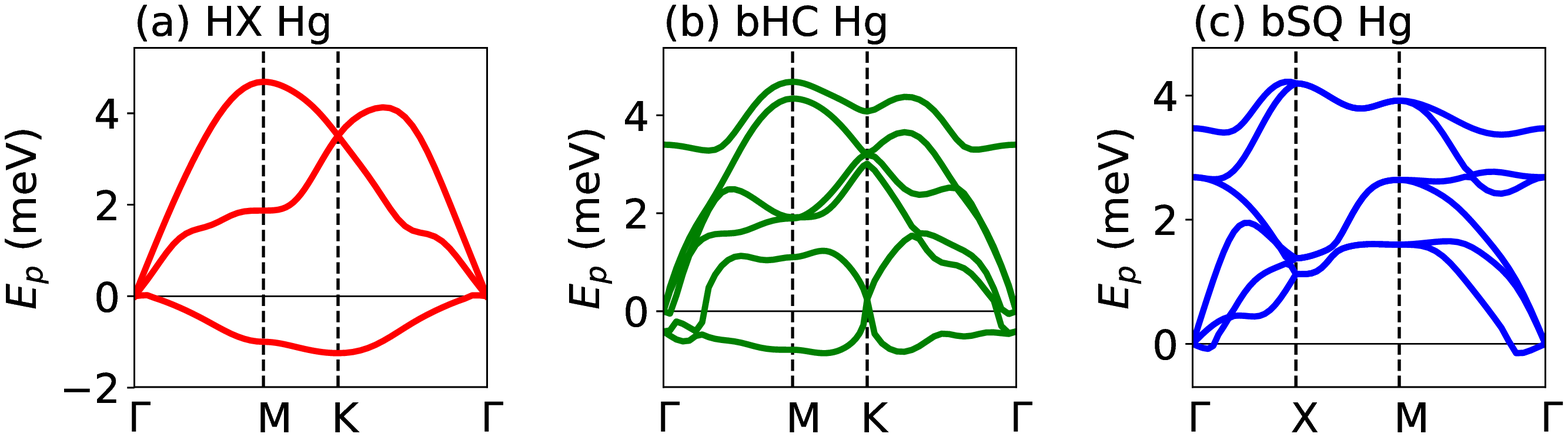}
\caption{\label{fig_Hg} Same as Fig.~\ref{fig_Li} but for Hg.}
\end{figure}

Figures \ref{fig_Zn}, \ref{fig_Cd}, and \ref{fig_Hg} show the phonon band structures of Zn, Cd, and Hg, respectively. For Zn and Cd, the HX and bHC structures are dynamically stable, while for Hg the bSQ structure is dynamically stable. 

\subsection*{Group 13: Al, Ga, In, and Tl}

\begin{figure}[H]
\center
\includegraphics[scale=0.30]{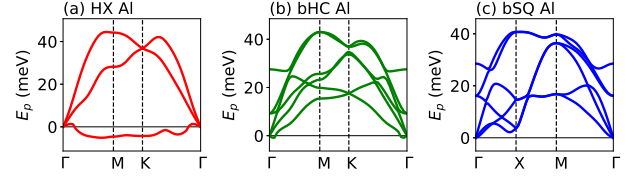}
\caption{\label{fig_Al} Same as Fig.~\ref{fig_Li} but for Al.}
\end{figure}

\begin{figure}[H]
\center
\includegraphics[scale=0.30]{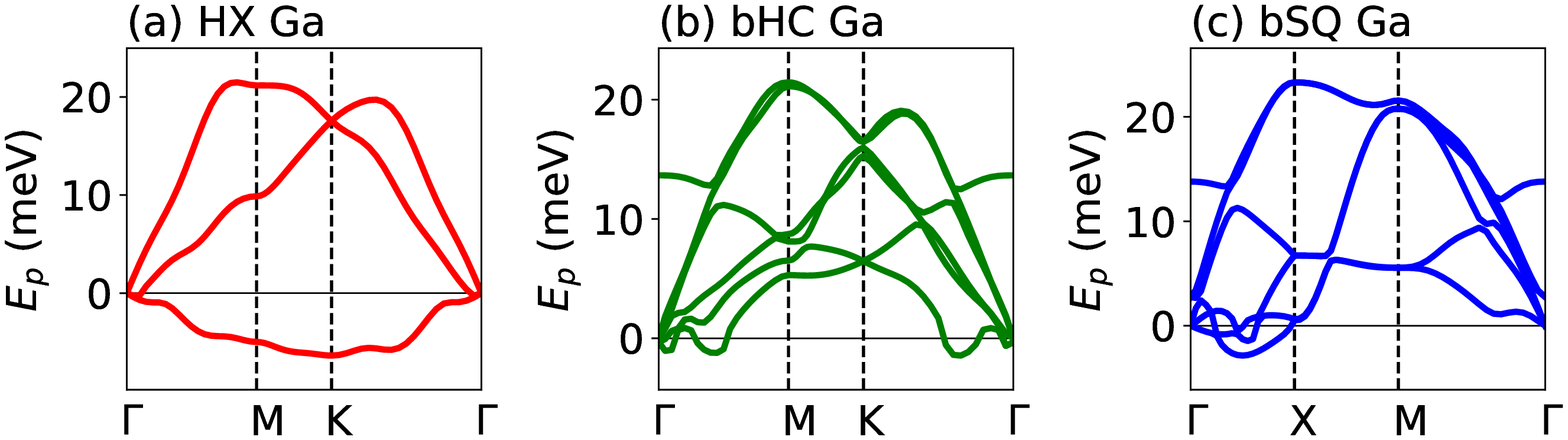}
\caption{\label{fig_Ga} Same as Fig.~\ref{fig_Li} but for Ga.}
\end{figure}

\begin{figure}[H]
\center
\includegraphics[scale=0.30]{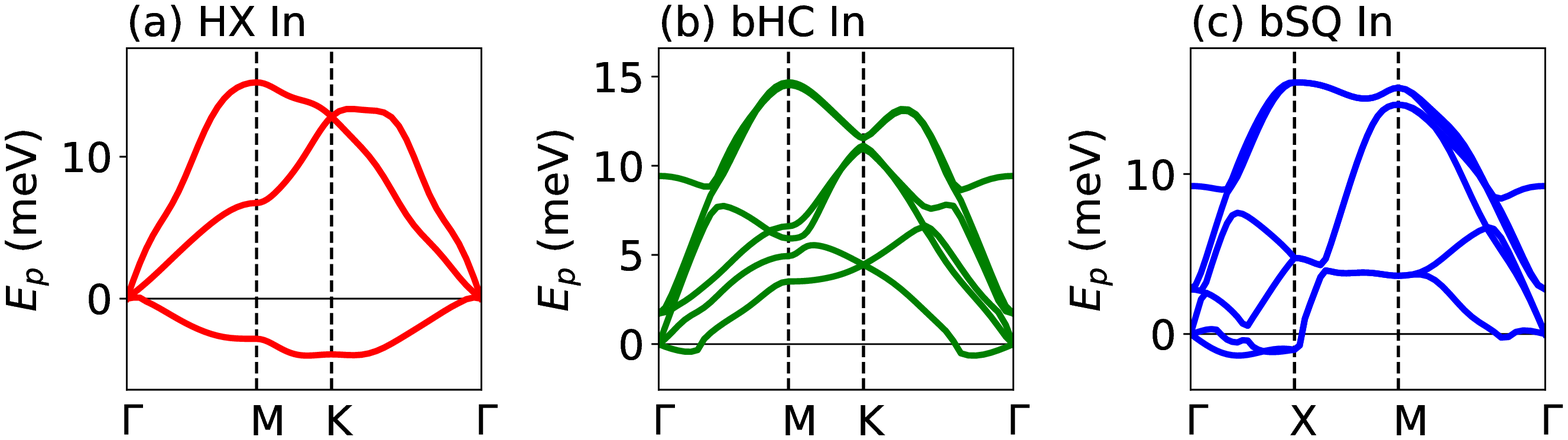}
\caption{\label{fig_In} Same as Fig.~\ref{fig_Li} but for In.}
\end{figure}

\begin{figure}[H]
\center
\includegraphics[scale=0.30]{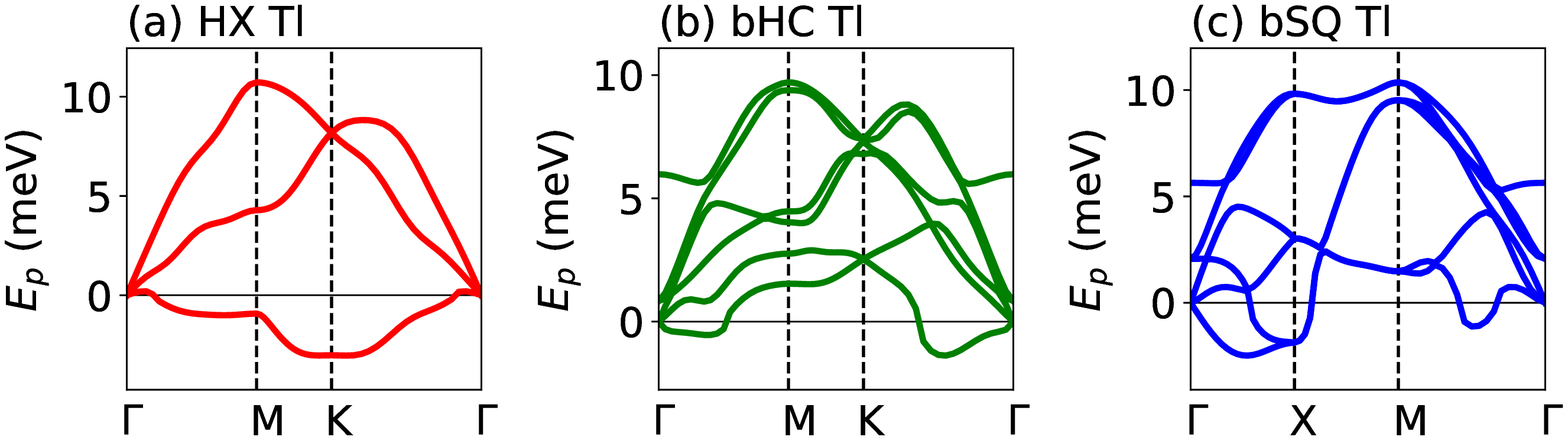}
\caption{\label{fig_Tl} Same as Fig.~\ref{fig_Li} but for Tl.}
\end{figure}

Figures \ref{fig_Al}, \ref{fig_Ga}, \ref{fig_In}, and \ref{fig_Tl} show the phonon band structure of Al, Ga, In, and Tl, respectively. For all the elements, the HX structure is unstable. The bHC and bSQ structures are dynamically stable in Al only. The stability of the trilayer structures will be investigated for Ga, In, and Tl, below. 

\subsection*{Group 14: Sn and Pb}

\begin{figure}[H]
\center
\includegraphics[scale=0.30]{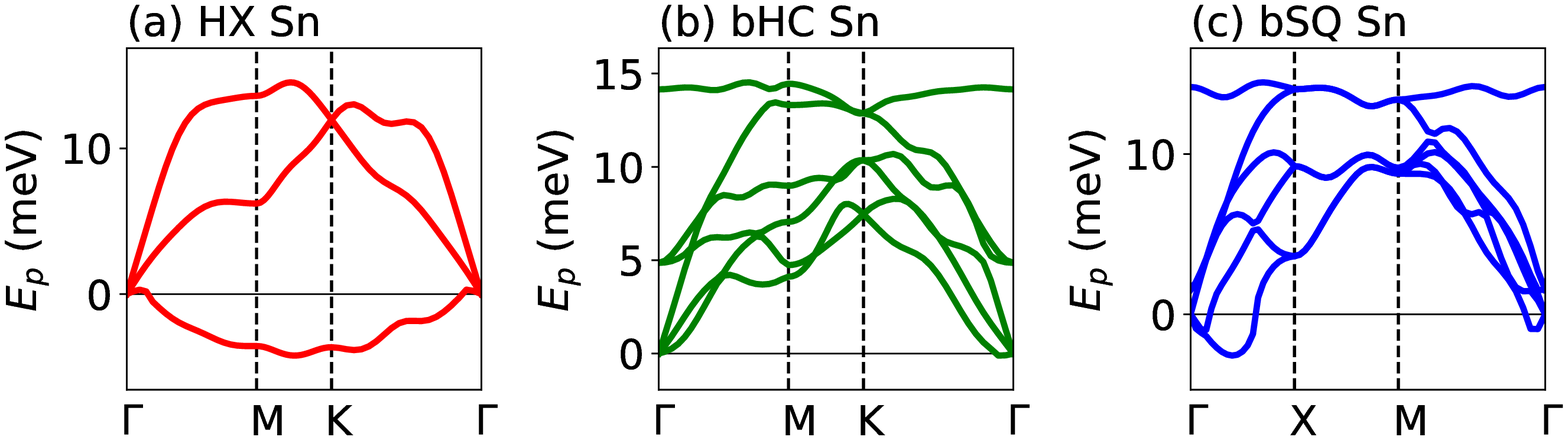}
\caption{\label{fig_Sn} Same as Fig.~\ref{fig_Li} but for Sn.}
\end{figure}

\begin{figure}[H]
\center
\includegraphics[scale=0.30]{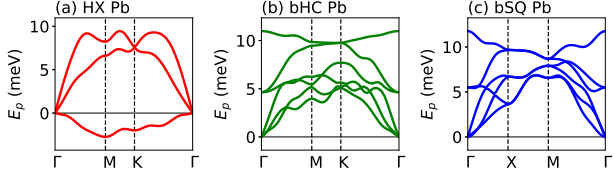}
\caption{\label{fig_Pb} Same as Fig.~\ref{fig_Li} but for Pb.}
\end{figure}

Figures \ref{fig_Sn} and \ref{fig_Pb} show the phonon band structure of Sn and Pb, respectively. Sn and Pb have the bHC as a dynamically stable structure. The bSQ Pb is a metastable phase. 

\subsection*{Spin-polarized Co and Ni}
\begin{figure}[H]
\center
\includegraphics[scale=0.4]{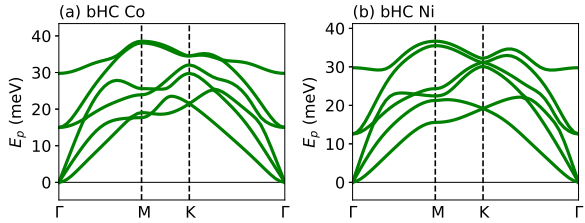}
\caption{\label{fig_spin} The phonon band structures of (a) bHC Co and (b) bHC Ni. Spin-polarized calculations are performed. }
\end{figure}

Figures \ref{fig_spin}(a) and \ref{fig_spin}(b) show the phonon band structures of bHC Co and bHC Ni, respectively. Both are dynamically stable. 

\subsection*{Trilayers}
\begin{figure}[H]
\center
\includegraphics[scale=0.40]{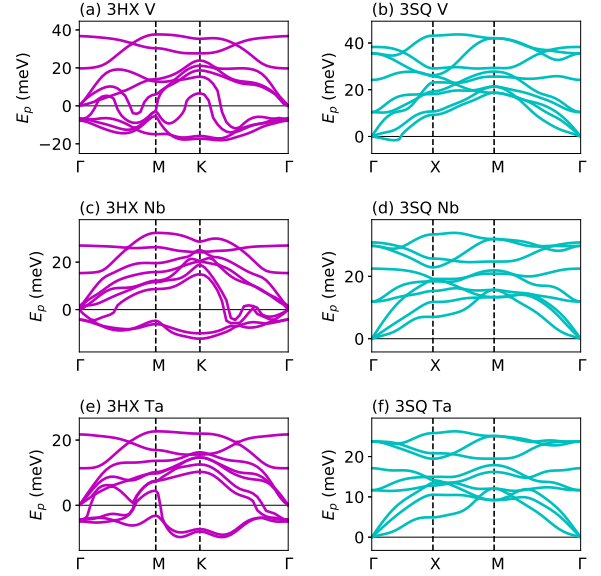}
\caption{\label{fig_tri_g5} The phonon band structures of group 5 metals of V, Nb, and Ta: Left for 3HX and right for 3SQ structures. }
\end{figure}

\begin{figure}[H]
\center
\includegraphics[scale=0.40]{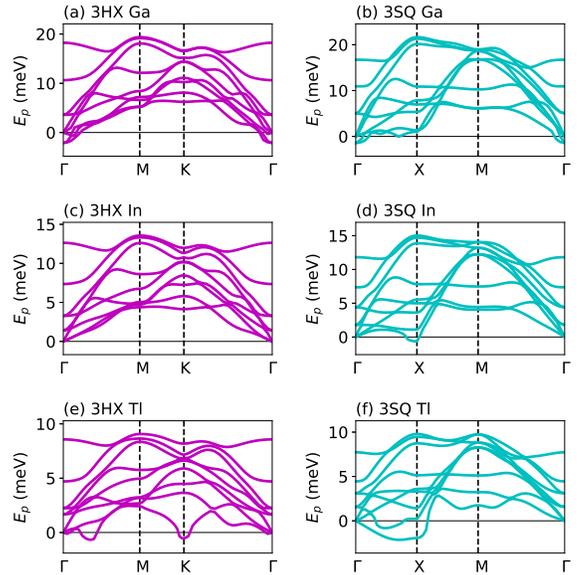}
\caption{\label{fig_tri_g13} Same as Fig.~\ref{fig_tri_g5} but for group 13 metals of Ga, In, and Tl. }
\end{figure}

Figures~\ref{fig_tri_g5} and \ref{fig_tri_g13} show the phonon band structures of group 5 metals (V, Nb, and Ta) and group 13 metals (Ga, In, and Tl), respectively: Left for 3HX and right for 3SQ structures. For group 5 metals, the 3SQ structure is dynamically stable, while the 3HX structure is unstable. For group 13 metals, the 3HX In is dynamically stable only. 


\subsection*{3D alloys}
\begin{figure}[H]
\center
\includegraphics[scale=0.35]{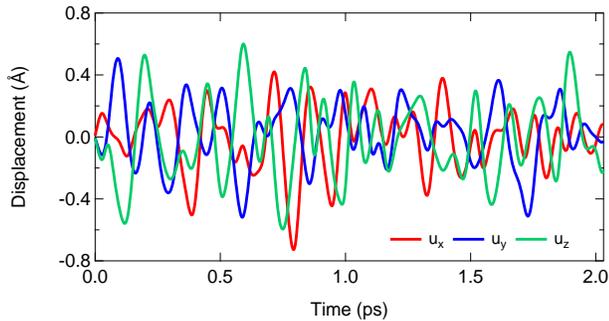}
\caption{\label{fig_md} The time-evolution of $\bm{u}$ for an Al atom in a unit cell of B$_h$ AlCu. }
\end{figure}

Figure \ref{fig_md} shows the time-evolution of the displacement vector $\bm{u}=(u_x,u_y,u_z)$ from the equilibrium position of an Al atom in a unit cell of B$_h$ AlCu at $T=2500$ K: The oscillation amplitude is about 0.4 \AA \ that is smaller than the lattice constant $a=2.787$ \AA. Above $T=3000$ K, no scf convergence is obtained after some MD steps, since the atoms deviate from the equilibrium position largely.

\begin{figure}[H]
\center
\includegraphics[scale=0.3]{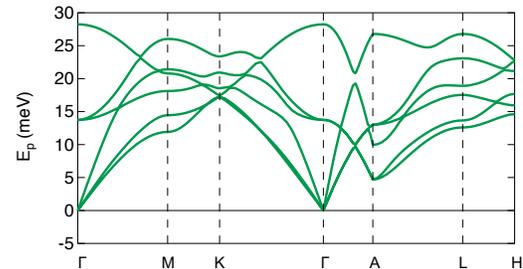}
\caption{\label{fig_cuzn} The phonon band structure of B$_h$ CuZn. }
\end{figure}

Figure \ref{fig_cuzn} shows the phonon band structure of B$_h$ CuZn. No imaginary phonon frequencies are observed, indicating that B$_h$ CuZn is dynamically stable at $T=0$ K. However, no scf convergence is obtained during the MD calculations assuming 2$\times$2$\times$2 supercell. 

\end{document}